\documentclass[preprint2,twocolumn,times,tighten]{aastex631}
\usepackage{subfigure}

\usepackage{newtxtext,newtxmath}
\usepackage{amsmath}
\usepackage{hyperref}
\usepackage{amssymb}
\usepackage{natbib}
\usepackage{morefloats}
\usepackage{multirow}
\usepackage{array}
\usepackage{verbatim}
\usepackage{color}

\usepackage{siunitx}

\begin{document}

\title{Anisotropic Velocity Fluctuations in Galaxy Mergers: A Probe of the Magnetic Field}

\email{yuehu@ias.edu; *NASA Hubble Fellow}

\author[0000-0002-8455-0805]{Yue Hu*}
\affiliation{Institute for Advanced Study, 1 Einstein Drive, Princeton, NJ 08540, USA }
\author{Joseph Whittingham}
\affiliation{Leibniz-Institute for Astrophysics Potsdam (AIP), An der Sternwarte 16, 14482 Potsdam, Germany}
\affiliation{Institut für Physik und Astronomie, Universität Potsdam, Karl-Liebknecht-Str. 24/25, 14476 Potsdam, Germany}
\author{Alex Lazarian}
\affiliation{Department of Astronomy, University of Wisconsin-Madison, Madison, WI, 53706, USA}
\author{Christoph Pfrommer}
\affiliation{Leibniz-Institute for Astrophysics Potsdam (AIP), An der Sternwarte 16, 14482 Potsdam, Germany}
\author{Siyao Xu}
\affiliation{Department of Physics, University of Florida, 2001 Museum Rd., Gainesville, FL 32611, USA}
\author{Thomas Berlok}
\affiliation{Niels Bohr Institute, University of Copenhagen, Blegdamsvej 17, 2100 Copenhagen, Denmark}


\begin{abstract}
Magnetic fields and turbulence are fundamental to the evolution of galaxies, yet their precise measurement and analysis present significant challenges. The recently developed Velocity Gradient Technique (VGT), which capitalizes on the anisotropy inherent in magnetohydrodynamic (MHD) turbulence, represents a new method for mapping magnetic fields in galaxies using spectroscopic observations. Most validations of VGT thus far, however, have relied upon idealized MHD turbulence simulations, which lack the more complex dynamics found in galaxies and galaxy mergers. In this study, we scrutinize VGT using an AREPO-based cosmological galaxy merger simulation, testing its effectiveness across pre-merger, merging, and post-merger stages. We examine the underlying assumptions of VGT and probe the statistics of gas density, velocity, and magnetic fields over time. We find that velocity fluctuations are indeed anisotropic at each stage, being larger in the direction perpendicular to the local magnetic field, as required by VGT. We find, additionally, that galaxy mergers substantially intensify velocity and density fluctuations and amplify magnetic fields at all scales. The observed scaling of the velocity fluctuations shows a steeper trend than $r^{1/2}$ between 0.6 and 3~kpc and a shallower trend at larger scales. The scaling of the magnetic field and density fluctuations at scales $\lesssim$ 1.0 kpc also predominantly aligns with $r^{1/2}$. Finally, we compare results from VGT to those derived from polarization-like mock magnetic field measurements, finding consistent and statistically significant global agreement in all cases. 
\end{abstract}



\section{Introduction}
Magnetic fields and turbulence are ubiquitous in the interstellar medium (ISM) and play a pivotal role in shaping astrophysical phenomena \citep{1995ApJ...443..209A, 2001SSRv...99..243B,2010ApJ...710..853C,Crutcher12,BG15,2017ApJ...835....2X,2020NatAs...4.1064H,2022ApJ...934....7H,2024MNRAS.527.3945H}. These forces exert influence over the acceleration and propagation of cosmic rays at multi-scales, thereby impacting ISM dynamics \citep{1949PhRv...75.1169F,1966ApJ...146..480J,2002PhRvL..89B1102Y,2004ApJ...614..757Y,2010A&A...510A.101F,2013ApJ...779..140X,2020ApJ...894...63X,2021MNRAS.501.4184H,2021arXiv211115066H,2022MNRAS.515.4229P}. On parsec scales, within the realms of molecular clouds, magnetic fields and turbulence are key in the star formation process
\citep{1965QJRAS...6..265M,MK04,2020MNRAS.499..597B,MO07,2012ApJ...757..154L,2012ApJ...761..156F,2018FrASS...5...39W,2020NatAs...4..704A,2021ApJ...912....2H,2022A&A...658A..90A,2023MNRAS.524.4431H,2023A&ARv..31....4R}. Recent studies have also shed light on their crucial function in directing gas flows to galactic nuclei, which is vital for the growth of supermassive black holes \citep{2012ApJ...751..124K,2013pss5.book..641B,2019MNRAS.489.3368B,2022ApJ...941...92H,2023MNRAS.526..224W}. Despite their critical importance, our comprehension of magnetic fields and turbulence in the ISM is still predominantly centered on our own Galaxy, with a broader understanding of their properties and impact across diverse galactic conditions remaining a poorly explored area.

Alongside traditional plane-of-the-sky (POS) magnetic field tracing methods such as polarized dust thermal emission \citep{Lazarian07,BG15,2015A&A...576A.104P,2020A&A...641A..11P,2016ApJ...824..134F,2021MNRAS.tmp.3119L,2022ApJ...936...65L,2023arXiv230904173L,2023MNRAS.519.3736H,2023arXiv231017048H} and polarized synchrotron radiation \citep{2001SSRv...99..243B,2002ARA&A..40..319C,2008A&A...482..783X,2011MNRAS.412.2396F,2015A&ARv..24....4B,2016A&A...594A..25P,ZhangLazarian2019,2021ApJ...920....6G}, the Velocity Gradient Technique (VGT; \citealt{GL17,LY18a,HYL18}) has been recently proposed as another approach for mapping magnetic fields in external galaxies \citep{2022ApJ...941...92H,2023MNRAS.519.1068L,2023ApJ...946....8T}. 

VGT is rooted in MHD turbulence \citep{GS95} and turbulent reconnection theory \citep{LV99}, which posit that MHD turbulence is characterized by anisotropic energy distribution and thus anisotropic turbulent eddies. These eddies are elongated along the local magnetic field, implying that velocity fluctuations are more pronounced perpendicular to the magnetic field \citep{2021ApJ...911...37H,2021ApJ...915...67H}. Consequently, the gradients of these velocity fluctuations, being steepest perpendicular to the magnetic field, act as effective detectors of the magnetic field's orientation.

The anisotropy of MHD turbulence, as predicted by theory, has been substantiated through numerical simulations \citep{CV20,2001ApJ...554.1175M,CL03,2010ApJ...720..742K,ZhangHCL2020,2021ApJ...910...88X,2021ApJ...911...37H} and in situ measurements of the solar wind \citep{2016ApJ...816...15W,2020FrASS...7...83M,2021ApJ...915L...8D,2023arXiv230512507Z}. However, the study of MHD turbulence in external galaxies remains limited. Particularly, understanding the development of both the turbulent anisotropy and the injection scale of turbulent energy is crucial for the effective application of VGT to other galaxies. 

Merger events in galaxies are of particular interest due to their unique dynamics. During a merger, tidal interactions can drive turbulence \citep{2013MNRAS.432.2529I} and amplify magnetic fields by the shearing and compression of gas \citep{2021MNRAS.506..229W}. The injection of turbulence may further amplify magnetic fields through the small-scale turbulent dynamo \citep{2001ApJ...550..752V,2009A&A...494...21A,2012PhRvL.108c5002B,2016ApJ...833..215X,2022MNRAS.515.4229P,2022ApJ...941..133H}. This can significantly impact the gas dynamics, cosmic ray acceleration, and star formation rates during a galaxy's evolution \citep{2006ApJS..163....1H,2020NatAs...4.1064H,2021MNRAS.506..229W}.

Studying MHD turbulence and magnetic fields in external galaxies -- even quiescent ones, however, has historically presented observational and numerical challenges. With the advent of the moving-mesh AREPO code \citep{2010MNRAS.401..791S,2016MNRAS.455.1134P,2020ApJS..248...32W}, amongst others, we may now study this physics in a cosmological context. In particular, it is possible to study magnetic fields and turbulence in high-resolution cosmological simulations of gas-rich mergers \citep{2021MNRAS.506..229W}. This development is crucial as it allows us to effectively study MHD turbulence and assess the VGT under realistic, physical conditions. We investigate this here for the first time, separating our analysis into pre-merger, merging, and post-merger phases. This choice encompasses the case of more isolated galaxies and dynamically interacting galaxies. As well as evaluating VGT, taking this strategy also allows us to study how MHD turbulence is driven by a merger event.

This paper is structured as follows: in  \S~\ref{sec:theory}, we briefly revisit the fundamental concepts of MHD turbulence anisotropy and how to obtain velocity information from spectroscopic observation. In \S~\ref{sec:data}, we describe the AREPO simulations we analyze and explain the process by which we transform these into mock observations. We also introduce the data analysis tools of structure-function and the VGT pipeline. In \S~\ref{sec:result}, we present our findings on the properties of gas density, velocity, and magnetic field fluctuations in pre-merger, merger, and post-merger galaxies. We also assess the effectiveness of VGT in tracing magnetic fields. In \S~\ref{sec:dis}, we delve into the implications of our findings for understanding MHD turbulence in external galaxies and the potential applications of VGT. Finally, we summarise our conclusions in \S~\ref{sec:con}.

\section{Theoretical considerations}
\label{sec:theory}
\subsection{Anisotropic MHD turbulence: the velocity gradient is statistically perpendicular to the local magnetic field}
\label{sec.2.1}
Though initially conceptualized as isotropic \citep{1963AZh....40..742I, 1965PhFl....8.1385K}, subsequent theoretical and numerical studies have established that MHD turbulence exhibits anisotropy 
\citep{1981PhFl...24..825M, 1983JPlPh..29..525S, 1984ApJ...285..109H, 1995ApJ...447..706M}. A pivotal model in understanding this anisotropy was introduced by \citet{GS95}, which centers around the concept of ``critical balance''. This principle posits that a balance exists between the cascading time of turbulence and the wave periods:
\begin{equation}
    (k_\bot v_l)^{-1}\sim(k_\parallel v_A)^{-1}.
\end{equation}
In this model, $k_\parallel$ and $k_\bot$ represent the components of the wavevector parallel and perpendicular to the magnetic field, respectively. The turbulent velocity at scale $l$ is further denoted by $v_l$ and $v_A$ represents the Alfv\'en speed. 

Considering that Kolmogorov-type turbulence adheres to the scaling relation $v_l \propto(l / L_{\rm inj})^{1/3}v_{\rm inj}$, where $v_{\rm inj}$ is the turbulent injection velocity at the injection scale, $L_{\rm inj}$, one may derive the anisotropy scaling:
\begin{equation}
\label{eq.gs95}
k_\parallel \propto k_\bot^{2/3}.
\end{equation}
This equation implies that as $k_\bot$ increases, $k_\parallel$ increases at a slower rate, indicating that the turbulent eddies become more elongated along the magnetic field lines as we go to smaller (physical) scales. This model of MHD turbulence has been crucial in enhancing our understanding of the complex interplay between magnetic fields and turbulent flows. In particular, it has the important implication that \textit{the elongation direction of turbulent eddies is indicative of the magnetic field.} 

It is important to note, however, that the utilization of wavevectors in Fourier space abstracts away the local spatial information. Consequently, the anisotropy described by Eq.~\ref{eq.gs95} is relative to the mean magnetic field, rather than local magnetic fields that penetrate the turbulent eddies at scale $l$. To trace the intricate magnetic field morphology within a galaxy, rather than a single mean magnetic field orientation, it is essential to consider the anisotropy of turbulent eddies in the frame of local magnetic fields. This anisotropy was derived by \citet{LV99}:
\begin{equation}
\label{eq.lv99}
 l_\parallel= L_{\rm inj}\left(\frac{l_\bot}{L_{\rm inj}}\right)^{2/3}M_A^{-4/3},~~~M_A\le 1,
\end{equation}
where $l_\bot$ and $l_\parallel$ represent the perpendicular and parallel scales of eddies relative to the local magnetic field, respectively, and $M_A = v_{\rm inj}/v_A$ is the Alfv\'en Mach number. 

To derive Eq.~\ref{eq.lv99}, \citet{LV99} considered the ``critical balance'' condition in the local reference frame: $v_{l,\bot}l_\bot^{-1}\sim v_Al_\parallel^{-1}$, assuming that the motion of eddies perpendicular to the local magnetic field direction obeys the Kolmogorov law in the strong turbulence regime: $ v_{l,\bot} = (l_\bot/L_{\rm inj})^{1/3} v_{\rm inj}M_A^{1/3}$ (see \citealt{xu2019study} for a review). This follows from the fact that, in the presence of fast turbulent reconnection, the motions of eddies perpendicular to the local direction of the magnetic field should experience only minimal resistance to the turbulent cascade. Consequently, \textit{turbulence predominantly cascades perpendicularly to the local magnetic field}. 

Combining Eq.~\ref{eq.lv99} with the ``critical balance'' condition yields the anisotropy scaling for the turbulent velocity:
\begin{equation}
\label{eq.v}
 v_{l,\bot}= v_{\rm inj}\left(\frac{l_\bot}{L_{\rm inj}}\right)^{\frac{1}{3}}M_A^{1/3},~~~ M_A\le 1,
\end{equation}
which has been demonstrated by numerical simulations \citep{CV20, 2001ApJ...554.1175M, CL03,  2010ApJ...720..742K, 2021ApJ...911...37H} and in situ measurements in the solar wind \citep{2016ApJ...816...15W, 2020FrASS...7...83M, 2021ApJ...915L...8D, 2023arXiv230512507Z}. Eq.~\ref{eq.lv99} further implies that $l_\parallel \gg l_\bot$. This means that the \textit{gradient of the turbulent velocity field is dominated by the component perpendicular to the local magnetic field}:
\begin{equation}
\label{eq.vg}
\nabla v_l\approx\frac{v_{l,\bot}}{l_{\bot}}=\frac{v_{\rm inj}}{L_{\rm inj}}M_A^{1/3}\left(\frac{l_{\bot}}{L_{\rm inj}}\right)^{-2/3},~ M_A \le 1.
\end{equation}
This scaling relation is particularly important because it shows that the gradient amplitude is proportional to $l_\bot^{-2/3}$ if the velocity fluctuations adhere to a Kolmogorov scaling. However, if the fluctuations follow, for example, a Burgers scaling characterized by $v_{l,\bot} \propto (l_\bot / L_{\rm inj})^{1/2} v_{\rm inj}$, the dependence of the gradient amplitude on the scale changes to  $l_\bot^{-1/2}$. Identifying the scaling of the velocity gradient can therefore inform us about the type of turbulence in the gas.

In both scenarios, this scaling indicates that the turbulent velocity gradient amplitude increases toward smaller scales—a behavior not typically seen in non-turbulent velocity fields (e.g., such as the galactic differential rotation), which are often scale-independent and become subdominant at smaller scales. The growth of velocity gradients at decreasing scales is intrinsic to turbulence, while the anisotropy is essential for aligning these gradients with the magnetic field. 

Finally, in order to apply VGT, it is clearly necessary that observations resolve the scale at which the anisotropy of the turbulence becomes significant. This scale is typically at $L_{\rm inj}$ for sub-Alfv\'enic and trans-Alfv\'enic turbulence \citep{xu2019study}. Such resolution also ensures that the observed velocity gradients are predominantly indicative of turbulent motions rather than larger-scale galactic dynamics. Early applications of the VGT, as conducted in studies by \cite{2022ApJ...941...92H,2023MNRAS.519.1068L}, have adopted an injection scale of about 100 pc, based on turbulence studies within the Milky Way \citep{1995ApJ...443..209A, 2010ApJ...710..853C,2022ApJ...934....7H}. We will examine $L_{\rm inj}$ in this study. 

\subsection{Effects of inflow and outflow: changing the gradient direction by 90 degrees} In \S\ref{sec.2.1}, we considered the dominance of MHD turbulence. However, in regions beyond the disk, outflows and inflows can become more prominent, affecting the alignment of the velocity gradient with the magnetic field. For example, when the thermal pressure or self-gravity is stronger than the magnetic and turbulent stresses, gas flows drag magnetic field lines with them. This may cause the acceleration of gas to be most marked parallel to those field lines. The result is that the orientation of the velocity gradient transitions to being parallel to the magnetic fields, rather than perpendicular to them\footnote{An exception to this behavior is an outflow that moves through a clumped medium such as the circumgalactic medium. In this case, magnetic fields can drape around the leading face of the dense obstacle and increase the magnetic energy density there until it balances the ram pressure seen by the obstacle \citep{2006MNRAS.373...73L,2008ApJ...677..993D,2010NatPh...6..520P}.}\citep{2020ApJ...897..123H}.

Outflows significantly alter fluid velocity statistics and affect velocity gradients in a manner analogous to inflows \citep{2022MNRAS.511..829H}. Additionally, however, outflows are typically expelled from the galactic center at high velocities but with lower velocities towards the periphery. As a result, outflows induce additional velocity gradients pointing from the outskirts to the center. This can also cause the orientation of velocity gradients to shift from being perpendicular to being aligned with the magnetic fields \citep{ 2024MNRAS.530.1066L}. 

\subsection{Velocity caustics: obtaining velocity information from spectroscopic observations}
In our analysis, we adopt the VGT methodology as employed in previous studies \citep[see, e.g.][]{2022ApJ...941...92H,2023MNRAS.519.1068L}. This approach is based on the anisotropy of turbulent velocity fluctuations, as described by Eqs.~\ref{eq.lv99}, \ref{eq.v}, and \ref{eq.vg}. The anisotropic nature of these velocity fields can be discerned from velocity channel maps, $p(x,y,v_{\rm los})$, obtained through spectroscopic observations, using the velocity caustics effect \citep{LP00,LP04,2016MNRAS.461.1227K,2023MNRAS.524.2994H}. 

The velocity caustics effect,  as first applied to the field of the interstellar medium by \citet{LP00}, highlights the distortion of density structures due to turbulent and shear velocities along the line-of-sight (LOS). Density structures possessing different velocities may be sampled into the same velocity channel, significantly modifying the observed intensity structures in the spectroscopic channel map. 

Mathematically, the density field $\rho(\pmb{x})$ can be decomposed into a mean density and zero-mean fluctuations: $\rho(\pmb{x}) = \langle\rho\rangle + \langle\rho \rangle\delta(\pmb{x})$, where $\pmb{x}=(x,y,z)$ represents spatial coordinates. The observed intensity in a velocity channel can then be represented as the sum of two terms \citep{2023MNRAS.524.2994H}, $p(x,y,v_{\rm los})=p_{vc}(x,y,v_{\rm los})+p_{dc}(x,y,v_{\rm los})$, in which:
\begin{align}
\label{eq:rhov}
p_{vc}(x,y,v_{\rm los})& \equiv \int_{v_{\rm los}-\Delta v/2}^{v_{\rm los}+\Delta v/2} \!\! dv \int \langle\rho\rangle \phi(v,\pmb{x}) dz, \\
\label{eq:rhod}
p_{dc}(x,y,v_{\rm los})& \equiv \int_{v_{\rm los}-\Delta v/2}^{v_{\rm los}+\Delta v/2} \!\! dv \int \langle\rho\rangle\delta(\pmb{x})\phi(v,\pmb{x}) dz,
\end{align}
which have units of mass density times length. Here, $\phi$ represents a Maxwellian distribution and the term $p_{vc}$ encompasses the mean intensity in the channel and carries fluctuations exclusively produced by velocity, the so-called \textit{velocity caustics} effect \citep{LP00} (see also \S~\ref{subsec:3.3}) \footnote{While the analytical theory by \cite{LP00} initially focused on the formation of velocity caustics in isothermal gas, subsequent studies have also confirmed the significance of velocity caustics in multi-phase environments \citep{2021ApJ...910..161Y,2023MNRAS.521..230H,2023MNRAS.524.2994H}.}. The $p_{dc}$ term, the \textit{density caustics} effect, captures inhomogeneities in the density field. The dominance of either $p_{vc}$ or $p_{dc}$ is dependent on the channel width, i.e., the velocity resolution of the observation. A narrower channel width enhances the contribution from $p_{vc}$,  thereby better capturing the velocity fluctuations. The typical threshold for this is when the channel width $\Delta v$ is less than the velocity dispersion $\sqrt{\delta v^2}$ of the turbulent eddies under study, that is, $\Delta v < \sqrt{\delta v^2}$. Observationally, the velocity dispersion is typically calculated from the dispersion of a velocity centroid map, i.e., a moment-one map. 

\section{Methodology}
\label{sec:data}
\subsection{Simulation setup}
In this work, we analyze the MHD cosmological zoom-in simulation ``1330-3M'', originally presented in \citet{2021MNRAS.506..229W}. This simulation focuses on a disc galaxy that undergoes a gas-rich major merger with another disc galaxy at $z\approx0.54$. The mass ratio of the merger is roughly 1:2 and the stellar mass of the main progenitor is approximately $6\times10^{11}\,M_\odot$ at the time of the merger. After the merger, the remnant is able to regrow in relative isolation, rebuilding a large gas and stellar disc by $z=0$. At this time, the remnant is Milky Way-like in mass and morphology. The merger scenario was selected from the original Illustris simulation \citep{2014MNRAS.444.1518V, 2014MNRAS.445..175G}, using selection criteria outlined in \citet{2016MNRAS.462.2418S} and \citet{2017MNRAS.470.3946S}. Zoom-in initial conditions were then created using a modified version of the \textsc{N-GenIC} code \citep{2014MNRAS.445..175G}. The cosmological simulation has a box side length of 75 co-moving Mpc $h^{-1}$. Cosmological parameters were taken from WMAP-9 \citep{2013ApJS..208...19H}, with the density parameters for matter, baryons, and a cosmological constant, respectively, being $\Omega_\text{m} = 0.2726$, $\Omega_\text{b} = 0.0456$, and $\Omega_\Lambda = 0.7274$, and Hubble's constant being $H_0 = 100 \,h~{\rm {km~s}^{-1}}~{\rm {Mpc}^{-1}} = 70.4$ km s$^{-1}$ Mpc$^{-1}$.

The simulation was performed using the Auriga galaxy formation model \citep{2017MNRAS.467..179G} and the moving-mesh code \textsc{AREPO} \citep{2010MNRAS.401..791S, 2016MNRAS.455.1134P, 2020ApJS..248...32W}. The dark matter mass resolution in the highest-resolution region, which encompasses the target halo and its immediate surroundings, is set to $m_\mathrm{DM} = 1.64 \times 10^5\; \mathrm{M}_\odot$. Gas cells within this region are allowed to refine and de-refine, such that their mass stays within a factor of two of  $2.74 \times 10^4\; \mathrm{M}_\odot$. The softening length is scaled with the mean radius of the gas cell, with a lower limit set to $30\,h^{-1}$ cpc. This value is co-moving until $z=1$, at which point it is frozen in physical units \citep[see sec 2.1 of ][for further details]{2021MNRAS.506..229W}. 

The ISM is described by the \citet{2003MNRAS.339..289S} model, in which the gas follows a stiff equation of state above the onset of thermal instability. This onset begins in our simulation at a gas density of $n_\text{SF} = 0.13\, \mathrm{cm}^{-3}$, at which point stars form in proportion to the local free-fall time. Stellar feedback is replicated through wind particles, which form in numbers reflecting the number of stars in the mass range of $8 -100\;\text{M}_\odot$. Wind particles are launched isotropically after creation and interact thereafter only gravitationally until they reach a cell with gas density $n < 0.05\, n_\text{SF}$ or exceed a maximum travel time of approximately 25 Myr. This model typically produces smooth, regular winds directed away from the galaxy at late times.

A supermassive black hole with mass $1.4\times10^5 \; \text{M}_\odot$ is seeded in the gas cell with the lowest gravitational potential once the corresponding friend-of-friends \citep{1985ApJ...292..371D} halo exceeds $7.1\times10^{10} \; \text{M}_\odot$. The black hole then accretes predominantly through an Eddington-limited Bondi-Hoyle-Lyttleton model \citep{1944MNRAS.104..273B, 1952MNRAS.112..195B}. Energy is injected isotropically into the nearest 512 gas cells to mimic quasar feedback, with the injection rate being set proportional to the accretion rate. Radio mode accretion and feedback is also implemented, although subdominant in this simulation \citep[see section 3.6 of ][for further details]{2023MNRAS.526..224W}.

Magnetic fields are implemented in the ideal MHD approximation \citep{2011MNRAS.418.1392P, 2013MNRAS.432..176P} using an HLLD Riemann solver \citep{2005JCoPh.208..315M} and a Powell 8-wave divergence cleaning scheme \citep{1999JCoPh.154..284P}. This scheme has been shown to be highly robust in ensuring the divergence criterion despite the highly dynamic nature of the scenario we simulate \citep[see, e.g., section 3.1.3 of][]{2021MNRAS.506..229W}. The magnetic field is seeded in the initial conditions at $z=127$ with a strength of $10^{-14}$~co-moving Gauss. This strength leads to magnetic fields that are dynamically irrelevant outside of collapsed halos. Within such halos, however, adiabatic compression and turbulence quickly act to bring the field into equipartition \citep{2021MNRAS.506..229W, 2022MNRAS.515.4229P, 2024MNRAS.528.2308P}. Once amplified, the magnetic field is able to have a sizeable dynamic impact. Indeed, hydrodynamical simulations form systematically different merger remnants compared to their MHD analogs \citep{2023MNRAS.526..224W}.

Whilst the saturation strength of the magnetic field over cosmic time is not a function of resolution, the rate of amplification is \citep{2017MNRAS.469.3185P}. It was shown in \citet{2021MNRAS.506..229W} and \citet{2023MNRAS.526..224W} that this is especially important in mergers, where the injection of turbulence is sudden, decaying over $\sim$~Gyr timescales thereafter. In order to amplify the magnetic field in such a timeframe, we must resolve a sufficient amplification rate, $\Gamma$. This is equivalent to requiring a sufficient Reynolds number, $\mathrm{Re}$, as $\Gamma \propto \sqrt{\mathrm{Re}}$ \citep{2022MNRAS.515.4229P}. It is believed that simulating a magnetic dynamo requires resolving Reynolds numbers with $\mathrm{Re} > 100$ \citep{2012PhRvE..85b6303S, 2022MNRAS.513.2457K}. Such values are routinely reached in the Auriga model at our gas resolution and halo mass \citep{2024MNRAS.528.2308P}. Moreover, it was independently estimated by \citet{2023arXiv231017036K} that the simulation analyzed in this paper exhibits Reynolds numbers of $\mathrm{Re} \gtrsim1000$ in the inner 5 kpc during the merger. This is more than sufficient to amplify the magnetic field in the given timeframe. Power spectra presented in \citet{2021MNRAS.506..229W} show that yet higher resolution is unlikely to significantly increase the amount by which the magnetic field is amplified, as we already achieve a similar saturation strength compared to that observed in turbulent boxes \citep[see, e.g.,][]{2016JPlPh..82f5301F}. We therefore consider both the turbulence and the magnetic field values in this simulation to be sufficiently well-resolved for the needs of the following analysis.

\subsection{Uniform-grid smoothing}
In order to generate mock observations, we convert the AREPO simulations into uniform-grid cubes. This smoothing process involves assigning a smoothing length to each AREPO gas cell $i$ using the formula:
\begin{equation}
    h_{\rm sml}\coloneq\alpha V_i^{1/3},
\end{equation}
where $\alpha=2$ is a constant factor determined as optimal by tests conducted using the $yt$ software \citep{Turk2011}, and $V_i$ represents the volume occupied by the $i$th AREPO gas cell. In this process, we iteratively loop over each AREPO gas cell, systematically smoothing properties such as density, velocity, or magnetic field onto a $512^3$ uniform grid. These cubes are centered around the galaxy's core, covering a radius of 20 kpc. The resultant voxels have a side-length of 78~pc, approximately matching the resolution of current high-resolution spectroscopic observation of nearby galaxies \citep[see, e.g.,][]{2021ApJS..257...43L,2022MNRAS.512.1522D}.

\subsection{Mock spectroscopic observations}
\label{subsec:3.3}
To convert the now uniform data into a spectroscopic cube, we utilize the density field $\rho(\pmb{x})$ and the velocity field $\pmb{v}(\pmb{x})$. In Position-Position-Velocity (PPV) space, the intensity distribution of a spectral line is influenced by both the gas density and its velocity distribution along the LOS. The LOS velocity component $v_{\rm los}(\pmb{x})$ comprises of the turbulent velocity $v_{\rm tur}(\pmb{x})$, the coherent velocity shear $v_{\rm gal}(\pmb{x})$ (e.g., from galactic rotation), and a residual component due to thermal motions. This thermal component, represented as $v_{\rm th}(\pmb{x})=v_{\rm los}(\pmb{x})-v_{\rm tur}(\pmb{x})-v_{\rm gal}(\pmb{x})$, follows a Maxwellian distribution $\phi(v_{\rm los},\pmb{x})$, leading to atomic/molecular emission intensity fluctuations in PPV space. The PPV emission intensity $\rho_s(x,y,v_{\rm los})$ is then given by:
\begin{align}
\label{eq.max}
\rho_s(x,y,v_{\rm los})&=\kappa \int \rho(\pmb{x}) \phi(v_{\rm los},\pmb{x}) dz,\\
\phi(v_{\rm los},\pmb{x}) & \equiv \frac{1}{\sqrt{2\pi c_s^2}}\exp\left[-\frac{[v_{\rm los}-v_{\rm tur}(\pmb{x})-v_{\rm gal}(\pmb{x})]^2}{2c_s^2}\right],
\end{align}
where $\kappa$ is a constant relating the number of emitters to the observed intensities such that $\rho_s$ has the unit of mass density divided by time. $c_s=\sqrt{\gamma k_{\rm B}T/m}$ is the speed of sound. Here $m$ is the mean molecular mass of the atoms or molecules, $\gamma$ is the adiabatic index, $k_{\rm B}$ is the Boltzmann constant, and $T$ is the temperature, which varies if the emitter is not isothermal. 

By integrating $\rho_s(x,y,v_{\rm los})$ over a specific velocity range, $\Delta v$, known as the channel width,  we obtain a spectroscopic velocity channel:
\begin{equation}
\label{eq.p}
p(x,y,v_{\rm los})=\int_{v_{\rm los}-\Delta v/2}^{v_{\rm los}+\Delta v/2}\rho_s(x,y,v)dv.
\end{equation}
In turn, by varying $v_{\rm los}$ and collecting $p(x,y,v_{\rm los})$, we obtain a synthetic spectroscopic cube, on which we may perform our VGT analysis. 

\subsection{VGT pipeline}
\label{sec:vgt}
The VGT pipeline \citep{2022ApJ...941...92H,2023MNRAS.519.1068L} is summarized below: 

\textbf{ Step 1.} Each thin channel map $p(x, y, v_{\rm los})$ is convolved with the 3$\times$3 Sobel kernels $G_x$ and $G_y$ \citep{Sobel1990AnI3}:
\begin{equation}
    \begin{aligned}
        \nabla_{x} p(x, y, v_{\rm los}) = G_x * p(x, y, v_{\rm los}), \\
        \nabla_{y} p(x, y, v_{\rm los}) = G_y * p(x, y, v_{\rm los}),
    \end{aligned}
\end{equation}
where the asterisks denote convolutions, and $\nabla_{x} p(x, y, v_{\rm los})$ and $\nabla_{y} p(x, y, v_{\rm los})$ are the gradient components in the thin channel maps along the $x$ and $y$ axis. These are used to calculate the overall pixelized gradient map $\psi_g (x,y,v_{\rm los})$:
\begin{equation}
       \begin{aligned}
&\psi_g (x,y, v_{\rm los})=\tan^{-1}\left(\frac{\nabla_{y} p(x, y, v_{\rm los})}{\nabla_{x}p(x, y, v_{\rm los})}\right).
\end{aligned}
\end{equation}

\textbf{Step 2.} The resulting gradient map $\psi_g(x,y, v_{\rm los})$ is then processed with the sub-block averaging method \citep{2017ApJ...837L..24Y}\footnote{ Turbulence properties are inherently statistical. The anisotropy given by the critical balance condition is not necessarily realized in individual turbulence realizations. Therefore, averaging is required to reliably determine the magnetic field direction when using velocity gradients.}. This is divided into rectangular sub-blocks with a size of 16$\times$16 pixels. We have verified the effectiveness of this grid size in previous studies \citep{LY18a,2021ApJ...911...37H}. For each sub-block:
\begin{enumerate}
    \item a histogram of gradient orientation is produced and the histogram is fitted with a Gaussian distribution,
    \item the gradient orientation corresponding to the Gaussian distribution's peak is then taken as the most probable orientation of the gradient for that sub-block.
\end{enumerate}

\textbf{Step 3.} After the sub-block averaging, we obtain the averaged gradient angle map $\psi_{gs}(x,y, v_{\rm los})$ for the channel at velocity $v_{\rm los}$. Repeating steps 1 and 2 for every channel within the velocity range of interest, we can then construct the pseudo-Stokes parameters $Q_g(x,y)$ and $U_g(x,y)$ to mimic polarization measurement:
\begin{equation}
\label{eq.QU}
\begin{aligned}
    &Q_{\rm g} (x,y)  =  \int^{v_{\rm los, max}}_{v_{\rm los, min}} p(x, y, v_{\rm los}) \cos(2\psi_{gs}(x,y,v_{\rm los}))dv_{\rm los},\\
    &U_{\rm g} (x,y)  =  \int^{v_{\rm los, max}}_{v_{\rm los, min}} p(x, y, v_{\rm los}) \sin(2\psi_{gs}(x,y,v_{\rm los}))dv_{\rm los},
\end{aligned}
\end{equation}
where $v_{\rm los, max}$ and $v_{\rm los, min}$ represent the upper and lower levels of the velocity range used for integration.

The pseudo-Stokes parameters can give the information of velocity gradient orientation. As we expect velocity gradient is preferentially perpendicular to the magnetic field (see \S~\ref{sec:theory} and \citealt{2022ApJ...941...92H}). The POS magnetic field orientation can then be indirectly inferred from:
\begin{equation}
    \psi_{\rm B}(x,y) = \frac{1}{2}\tan^{-1}\left(\frac{U_{\rm g} (x,y) }{Q_{\rm g} (x,y) }\right) + \frac{\pi}{2}.
\end{equation}

\subsection{Mock polarization observations and Alignment Measure}
To validate the effectiveness of VGT in tracing the POS magnetic field orientation, we simulate Stokes $Q$ and $U$ maps through mock polarization observations. We do this by integrating along the LOS through the simulation cube, following the method in \cite{2015A&A...576A.105P}:
\begin{equation}
\label{eq.dust}
\begin{aligned}
Q(x,y)&\propto\int \rho\frac{B_x^2-B_y^2}{B^2}dz,\\
U(x,y)&\propto\int \rho\frac{2B_xB_y}{B^2}dz,\\
\tilde\psi_B(x,y)&=\frac{1}{2}\tan^{-1}\left(\frac{U}{Q}\right)+\frac{\pi}{2},
\end{aligned}
\end{equation}
where $\rho(\pmb{x})$ is gas mass density and $\tilde\psi_B$ is the magnetic field angle. $B(\pmb{x})$ denotes total magnetic field strength, while $B_x(\pmb{x})$ and $B_y(\pmb{x})$ are its $x$-axis component and $y$-axis component. Note that we do not fully model dust or synchrotron polarization here. Because VGT measures the gas density weighted magnetic field it is more appropriate to use gas density-weighted Stoke parameters for a comparison. 

\begin{figure*}
	\includegraphics[width=1.0\linewidth]{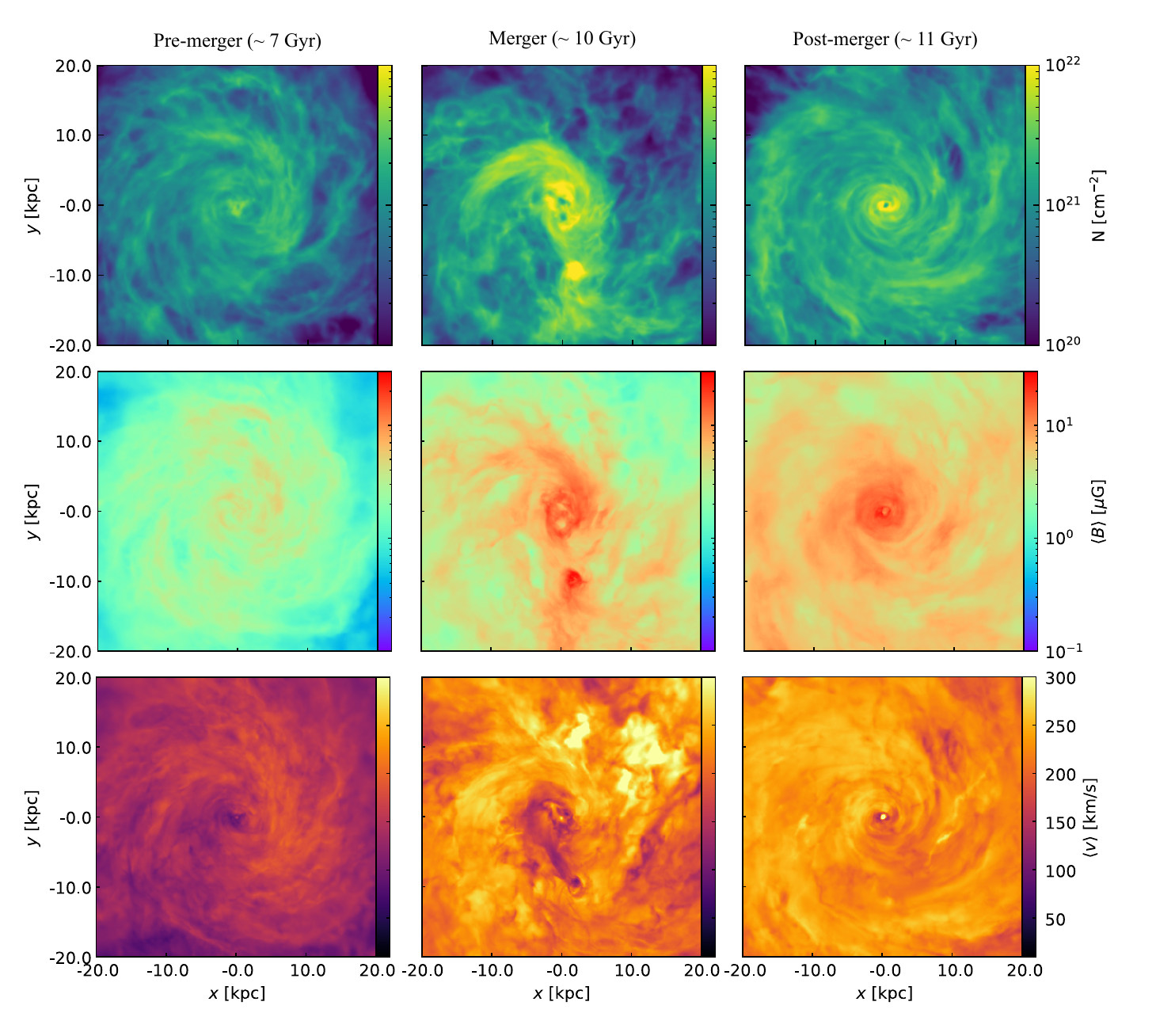}
    \caption{Maps of gas column density (1st row), mean volume-weighted magnetic field strength (2nd row), and the mean mass-weighted magnitude of the velocity (3rd row). 
    The galaxies are seen face-on and the depth of the LOS is 40~kpc. The 1st column corresponds to the pre-merger stage, while the 2nd column and 3rd column are associated with the merger and post-merger stages, respectively.}
    \label{fig:2Dmap}
\end{figure*}
To quantify the agreement between VGT and the magnetic field inferred from polarization, we utilize the Alignment Measure (AM; \citealt{GL17}), expressed as:
\begin{equation}
\begin{aligned}
\label{eq.am}
{\rm AM} = 2 \left(\cos^2\theta_{\rm r} - \frac{1}{2}\right).
\end{aligned}
\end{equation}
here, $\theta_{\rm r} = \vert\psi_{\rm B}-\tilde\psi_{\rm B}\vert$. An AM value of 1 implies parallel alignment of $\phi_{\rm B}$ and $\psi_{\rm B}$, while -1 indicates perpendicularity.

\begin{figure*}
	\includegraphics[width=1.0\linewidth]{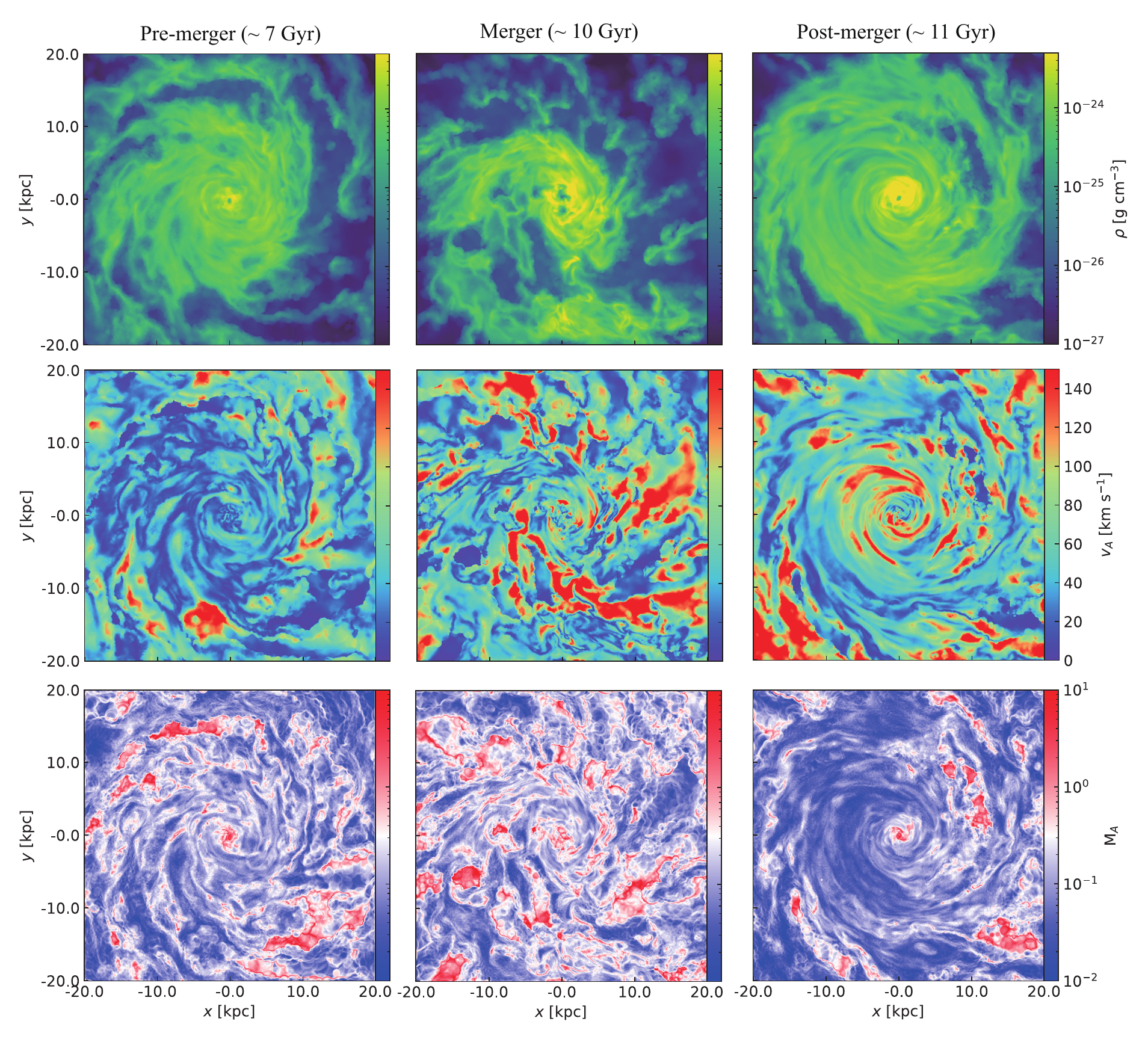}
    \caption{2D slice of gas mass density (1st row), Alfv\'en speed (2nd row), and pixel-scale Alfv\'en Mach number (3rd row). To calculate the Alfv\'en Mach number, we determine both the local Alfv\'en speed and the local velocity fluctuation at every pixel. The velocity fluctuation is obtained from the structure function (Eq.~\ref{eq.sf}) at a separation of one pixel ($\sim78$~pc).  The galaxies are viewed face-on, and the slices are extracted from the galaxy centers with a scale height of $\sim$100~pc. The 1st column corresponds to the pre-merger stage, while the 2nd column and 3rd column are associated with the merger and post-merger stages, respectively.}
    \label{fig:2Dslice}
\end{figure*}
\subsection{The second-order structure function}
Finally, we also utilize the second-order structure function in this study. This is a widely accepted method for quantifying the statistical properties of turbulent flows \citep{CV20,2021ApJ...910...88X,2021ApJ...911...37H}. The structure function for velocity is defined as:
\begin{equation}
\label{eq.sf}
{\rm SF_2}(r)\coloneq\langle|\pmb{v}(\pmb{r}_1)-\pmb{v}(\pmb{r}_2)|^2\rangle_r,
\end{equation}
where $\pmb{v}(\pmb{r}_1)$ and $\pmb{v}(\pmb{r}_2)$ are the velocities at position $\pmb{r}_1$ and $\pmb{r}_2$, and $r=|\pmb{r}_1-\pmb{r}_2|$ represents the separation. This function can be adapted to measure the magnetic field or gas density fluctuations by replacing $\pmb{v}(\pmb{x})$ with $\pmb{B}(\pmb{x})$ or $\rho(\pmb{x})$, respectively.

To explore anisotropy in turbulence, we decompose the structure-function into components parallel and perpendicular to the local magnetic fields $\pmb{B}_{\rm loc}$, following the method described in \citet{CV20}:
\begin{equation}
\label{eq.sf_loc}
\begin{aligned}
&\pmb{B}_{\rm loc}\coloneq\frac{1}{2}(\boldsymbol{B}(\pmb{r}_1)+\pmb{B}(\pmb{r}_2)),\\
&{\rm SF}_2^{v}(l_\bot,l_\parallel)=\langle|\pmb{v}(\pmb{r}_1)-\pmb{v}(\pmb{r}_2)|^2\rangle_r,
\end{aligned}
\end{equation}
where $r=\sqrt{l_\bot^2+l_\parallel^2}$ is defined in a cylindrical coordinate system, with the $\hat{\pmb{l}}_\parallel=\hat{\pmb{B}}_{\rm loc}$ and $l_\bot=|\hat{\pmb{l}}_\parallel\times\pmb{r}|$.

From these, we can derive the velocity fluctuations that are parallel ($\delta v_\parallel$) or perpendicular ($\delta v_\bot$) to the local magnetic fields as follows:
\begin{equation}
\label{eq.sf_v}
\begin{aligned}
\delta v_{\bot}&=\sqrt{{\rm SF}_2^v(l_\bot,0)},\\
\delta v_{\parallel}&=\sqrt{{\rm SF}_2^v(0,l_\parallel)}.
\end{aligned}
\end{equation}
This approach allows us to connect the properties of turbulent velocity fields with the magnetic fields.

\section{Results}
\label{sec:result}
\subsection{Evolution of column density, magnetic field, and velocity: from low pre-merger to peak, and high post-merger stages}
In Fig.~\ref{fig:2Dmap}, we present detailed maps showcasing the column density, mean volume-weighted magnetic field, mean mass-weighted velocity, and mean mass-weighted pixel-scale Alfv\'en Mach number, averaged along the LOS, focusing on a volume with side-length $\pm$20~kpc centered on the main progenitor. For the column density maps, a noticeable distortion in shape is observed during the merger when compared to the pre-merger view. This is accompanied by an increase in the central density by a factor of $3 - 5$ relative to before and after the merger. Gas accretion leads to a higher comparative average gas density in the galaxy post-merger.

\begin{figure}
\centering
\includegraphics[width=1.0\linewidth]{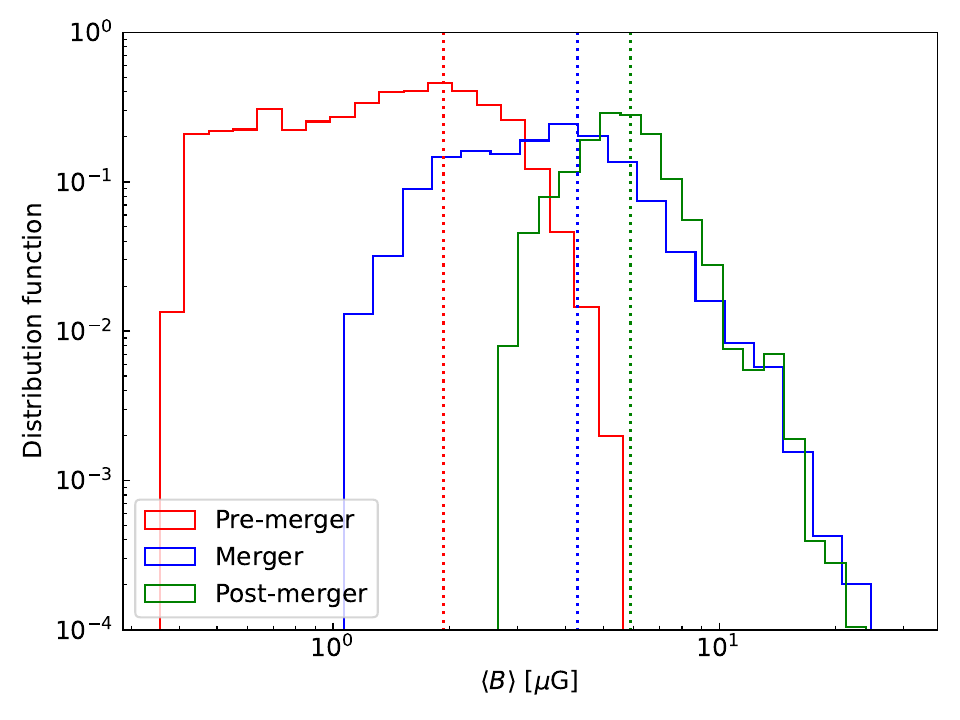}
        \caption{Histograms of mean volume-weighted magnetic field strength along the LOS, where data is taken from Fig.~\ref{fig:2Dmap}. The dashed lines represent each distribution's median value. The area beneath the curve is normalized to unity.
        }
    \label{fig:Bhist}
\end{figure}

We observe distinct strengths across different merger stages in examining the magnetic field averaged along the LOS. As the 2D maps in Fig.~\ref{fig:2Dmap} and histograms in Fig.~\ref{fig:Bhist} show, initially, at the pre-merger stage, the magnetic field strength varies between $\sim\SI{0.35}{\micro G}$ and $\sim\SI{5.62}{\micro G}$, with a median value of $\sim\SI{1.93}{\micro G}$. This value experiences a significant increase during the merger stage, ranging from $\sim\SI{1.06}{\micro G}$ to $\sim\SI{29.41}{\micro G}$, and the median value rising to $\sim\SI{4.30}{\micro G}$. The magnetic field strength undergoes yet further amplification in the post-merger stage, reaching levels between $\sim\SI{2.67}{\micro G}$ and $\sim\SI{27.07}{\micro G}$, with the median soaring to $\sim\SI{5.87}{\micro G}$. This amplification cannot be accounted for by gas compression alone and is, at least in part, driven by a turbulent dynamo mechanism \citep{2021MNRAS.506..229W}, which is on top of the small-scale dynamo that is driven by the gravitational collapse of the proto-galaxy and amplifies most of the galactic magnetic field \citep{2017MNRAS.469.3185P,2024MNRAS.528.2308P,2022MNRAS.513.6028L}. The velocity maps corroborate this finding, as there is a noticeable increase in velocity fluctuations in the post-merger stage compared to the pre-merger stage.

In Fig.~\ref{fig:2Dslice} presents 2D slices of gas mass density, Alfv\'en speed, and pixel-scale Alfv\'en Mach number. The slices are taken at the galaxy center with a scale height of approximately 100~pc. At the pre-merger stage, the gas mass density ranges from approximately $10^{-27}$ to $10^{-24}$~g~cm$^{-3}$. During the later merger and post-merger stages, the density at the galactic center increases by roughly one order of magnitude. Low-density regions ($<10^{-25}$~g~cm$^{-3}$) are typically associated with high Alfv\'en speeds, exceeding 40~km~s$^{-1}$ and sometimes reaching values above 150~km~s$^{-1}$. As a result, these regions are characterized by low $M_A$, indicating sub-Alfv\'enic conditions on a scale of 78~pc. However, localized regions with super-Alfv\'enic conditions are also present.

\begin{figure}
\centering
\includegraphics[width=1.0\linewidth]{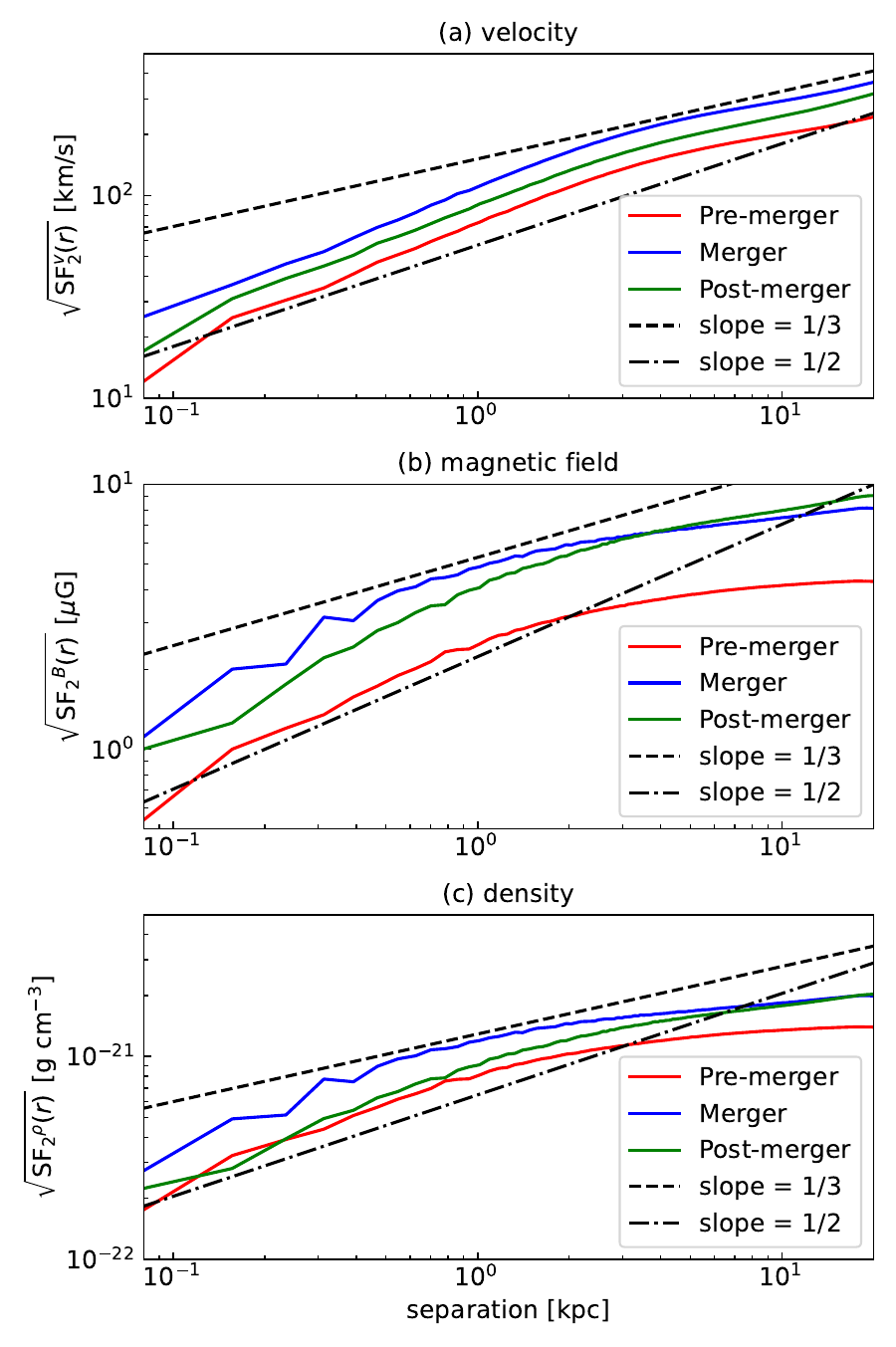}
        \caption{The square root of the second order structure function (see Eq.~\ref{eq.sf}). The structure function is calculated for gas velocity (top), magnetic field (middle), and gas density (bottom). To guide the eye, the dashed and dash-dotted lines represent power-law slopes of 1/3 and 1/2, for comparison with a Kolmogorov and Burgers scaling of turbulence, respectively.}
    \label{fig:SF}
\end{figure}

\begin{figure*}
\includegraphics[width=0.95\linewidth]{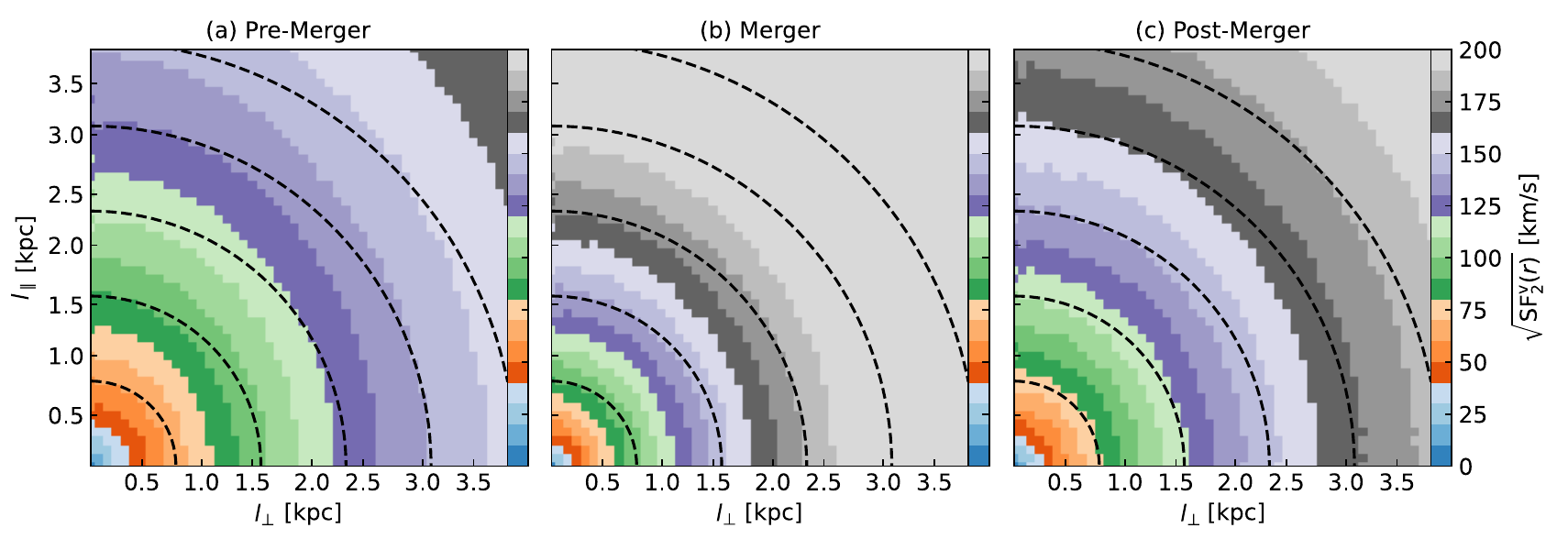}
        \caption{Maps of the velocity structure-function decomposed along the directions parallel ($l_\parallel$) and perpendicular ($l_\bot$) to the local magnetic field. The reference dashed lines represent isotropic cases.}
    \label{fig:SFv_decomposed}
\end{figure*}


\begin{figure}
    \centering
    \begin{subfigure}   
    \centering
    \includegraphics[width=0.99\linewidth]{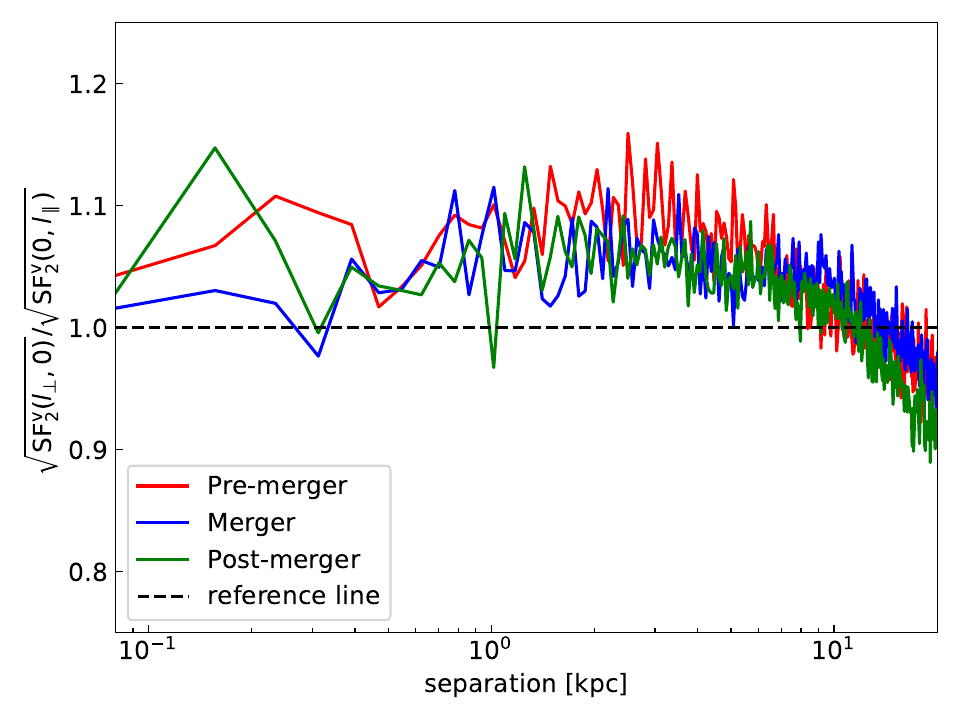}
   \end{subfigure}
    \centering
    \centering
    \begin{subfigure}   
    \centering
    \includegraphics[width=0.99\linewidth]{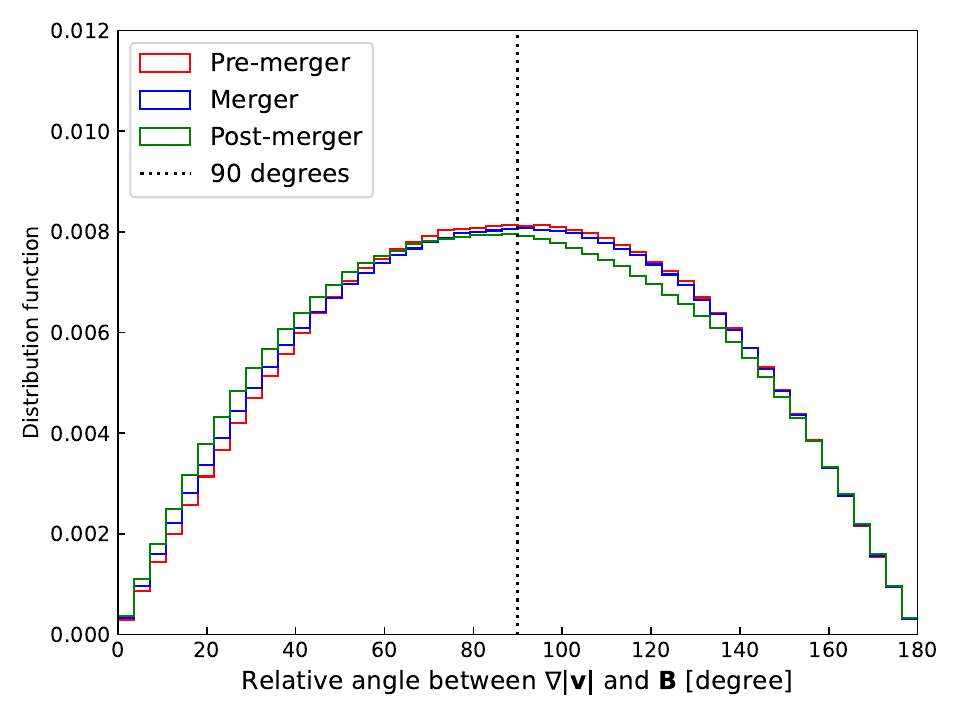}
   \end{subfigure}
     \begin{subfigure}
     \centering
    \includegraphics[width=0.99\linewidth]{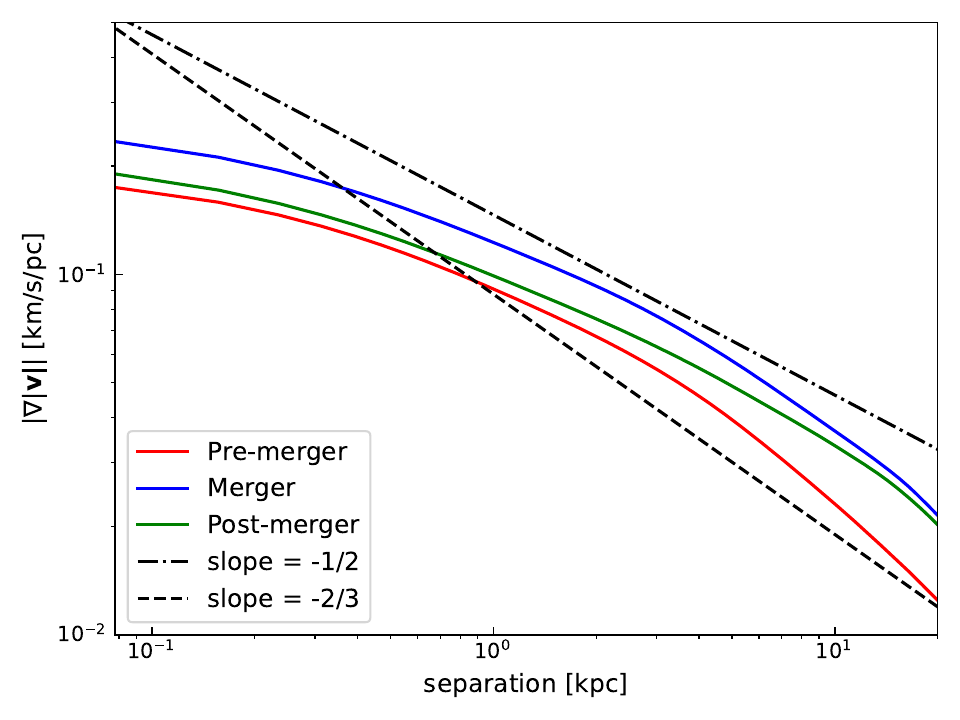}
   \end{subfigure}
    \caption{Top: The ratio of the decomposed structure functions, along the directions parallel ($0, l_\parallel$) and perpendicular ($l_\bot,0$) to the local magnetic field, as a function of separation. The reference dashed line represents the isotropic case. 
    Middle: Histogram of the relative angle between velocity gradient $\nabla |\pmb{v}|$ and $\pmb{B}$. The area beneath the curve is normalized to unity. Bottom: The velocity gradient amplitude as a function of separation scales. The dashed and dash-dotted lines have power-law slopes of -2/3 and -1/2: the expected scaling for the gradient of Kolmogorov turbulence and Burgers turbulence, respectively. }
    \label{fig:Bvangle}
\end{figure}

\subsection{Structure function analysis: velocity, magnetic field, and density fluctuations}
\label{sec:structure-functions}
Fig.~\ref{fig:SF} shows the square root of the second-order structure function (as defined in Eq.~\ref{eq.sf}) for velocity, magnetic field, and gas density. This measure effectively captures the fluctuation levels seen in Fig.~\ref{fig:2Dmap}. In the context of velocity, a noticeable increase in fluctuations is evident at the merger stage, in contrast to the pre-merger stage. This increase aligns with expectations, as galaxy mergers are expected to drive turbulence and amplify velocity fluctuations. Such fluctuations primarily arise from gravitational interactions, although stellar winds are also able to launch outflows, which generate turbulence in the circumgalactic medium (CGM) and at the ISM/CGM border \citep{2021MNRAS.501.4888V}. Such inflows and outflows are able to have a sizeable influence on the development of velocity fluctuations \citep{2022MNRAS.513.2100H,2024MNRAS.530.3431V}. 
At the post-merger stage, there is a decrease in velocity fluctuations, although their magnitude remains higher compared to the pre-merger stage. There are multiple possible origins for what sustains these velocity fluctuations, including a higher gas density, thicker disk, increased galactic nuclear activity, and/or formation of new spiral arms. All of these factors likely contribute at least to some extent.

During all three stages, The velocity structure function closely follows a slope of approximately 1/2 for separations up to about 0.3~kpc. At separations smaller than 0.15~kpc, the structure function steepens for both pre-merger and post-merger cases. However, a previous zoom-in analysis extending down to 0.03~kpc revealed that the scaling relation is consistent with a turbulent cascade at scales smaller than 0.1~kpc (see Fig.~15 in \citealt{2021MNRAS.506..229W}). Beyond 0.3~kpc, we observe a slightly steeper slope of $\sim$$1/2$ in the structure-function, before the slope becomes slightly shallower than $\sim$$1/3$ from $\gtrsim3$ kpc onwards. This scale is approximately the height of the thick disk. At these larger scales, the velocity fluctuations appear to be primarily governed by large-scale galactic motions rather than small-scale turbulent processes. 

Similar to the velocity structure functions, for scales smaller than approximately 3~kpc, the magnetic field structure function shows a slope steeper than the Kolmogorov slope of $\sim$$1/3$; indeed the slope is closer to a Burgers-style slope. Beyond 3~kpc, the slope becomes again much shallower than $\sim$$1/3$. However, the trend differs compared to the velocity structure function in the post-merger stage \footnote{It is important to note that velocity fluctuations are a more direct indicator of turbulence. Magnetic field and density fluctuations can exhibit different scalings, even when the velocity fluctuation follows a Kolmogorov-type or Burger-type spectrum \citep{2005ApJ...624L..93B,2007ApJ...658..423K,2024MNRAS.527.3945H}.}; unlike the decrease in velocity fluctuations here, the magnetic field fluctuations continue to exhibit a slight increase at scales larger than 3~kpc. This ongoing amplification can also be observed in Figs.~\ref{fig:2Dmap} and~\ref{fig:Bhist}. 

This trend can be understood by examining the final panel of Fig.~\ref{fig:SF}, in which we examine the structure functions for the gas density over time. These scalings show a similar evolution to those seen for the magnetic field. In particular, in addition to Burgers-type scalings below 3 kpc, it can be seen that fluctuations with separations $\gtrsim$5 kpc continue to be large post-merger, being at least comparable with the during-merger values. This implies that a large amount of the magnetic field amplification in the galaxy is due to adiabatic compression. The remaining difference between pre-merger and post-merger magnetic field strengths can be attributed to a small-scale dynamo \citep{2021MNRAS.506..229W}.

\subsection{Anisotropy of velocity fluctuations and velocity gradients}
To further understand the expected anisotropy inherent in MHD turbulence, we have undertaken a decomposition of the velocity structure functions (as outlined in Eq.~\ref{eq.sf_loc}) along axes parallel and perpendicular to the local magnetic fields. This result is presented in Fig.~\ref{fig:SFv_decomposed}, where the $l_\parallel$-axis corresponds to the direction parallel to the magnetic field, and the $l_\bot$-axis denotes the perpendicular orientation. Upon examination of the decomposed structure functions across all three stages of galactic evolution, a notable pattern emerges. The contours representing velocity fluctuations exhibit an elongation along the $l_\bot$-axis. This can be interpreted as follows: for any given separation scale, the magnitude of the velocity fluctuation is consistently greater in the direction perpendicular to the magnetic field (along the $l_\bot$-axis). This indicates that the velocity fluctuations exhibit anisotropy. 

In Fig.~\ref{fig:Bvangle}, we delve into a quantitative analysis of the velocity fluctuations, examining the ratio of the perpendicular and parallel structure function components (see Eq.~\ref{eq.sf_v}) as a function of scale. This ratio reveals that the magnitude of fluctuations occurring perpendicular to the magnetic fields is greater than the parallel component by a factor ranging between approximately 1.05 and 1.1. The magnitude of this ratio is correlated with the magnetization of the medium (see Eq.~\ref{eq.lv99}), with a more strongly magnetized medium having a larger anisotropy. The maximum ratio of around 1.1 suggests that the magnetic field energy and turbulent kinetic energy are approximately in equilibrium. 

This ratio remains relatively stable up to a scale of around 3~kpc. However, beyond this 3~kpc threshold, the ratio begins to decrease. Indeed, after a separation scale of 10 kpc, the ratio is less than unity. It is likely that at scales larger than 3~kpc, the velocity fluctuations start to be influenced by inflows and outflows \citep{2020ApJ...897..123H,2021ApJ...912....2H,2024MNRAS.530.1066L}. This influence eventually grows to the point that the parallel component dominates.



\begin{figure*}
\centering
\includegraphics[width=1.0\linewidth]{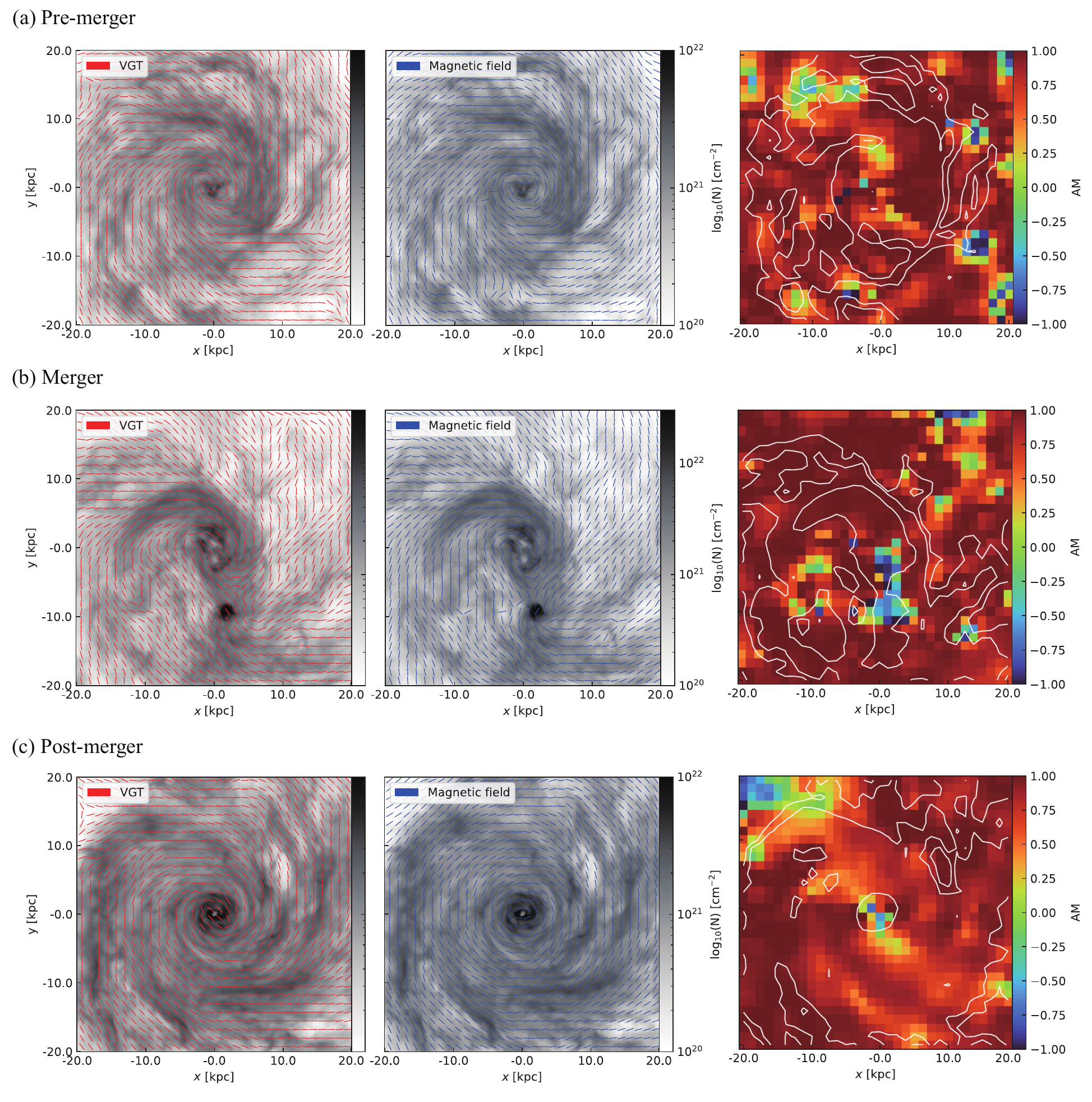}
        \caption{A comparison of the magnetic fields inferred from VGT (left, red line segment) and polarization (middle, blue line segment) at the pre-merger (top), merging (middle), and post-merger (bottom) stages. The magnetic field segments are overlaid with the gas column density map. The right panel shows the corresponding alignment measure (AM) maps, where a positive AM corresponds to parallel alignment between the two vectors, while a negative AM represents a perpendicular alignment, see Eq.~\ref{eq.am}. The contours shown in the AM maps represent the column density contours of $6\times10^{20}$, $10^{21}$, and $3\times10^{21}$ cm$^{-2}$.}
    \label{fig:vgt_060}
\end{figure*}

\begin{figure*}
\centering
\includegraphics[width=1.0\linewidth]{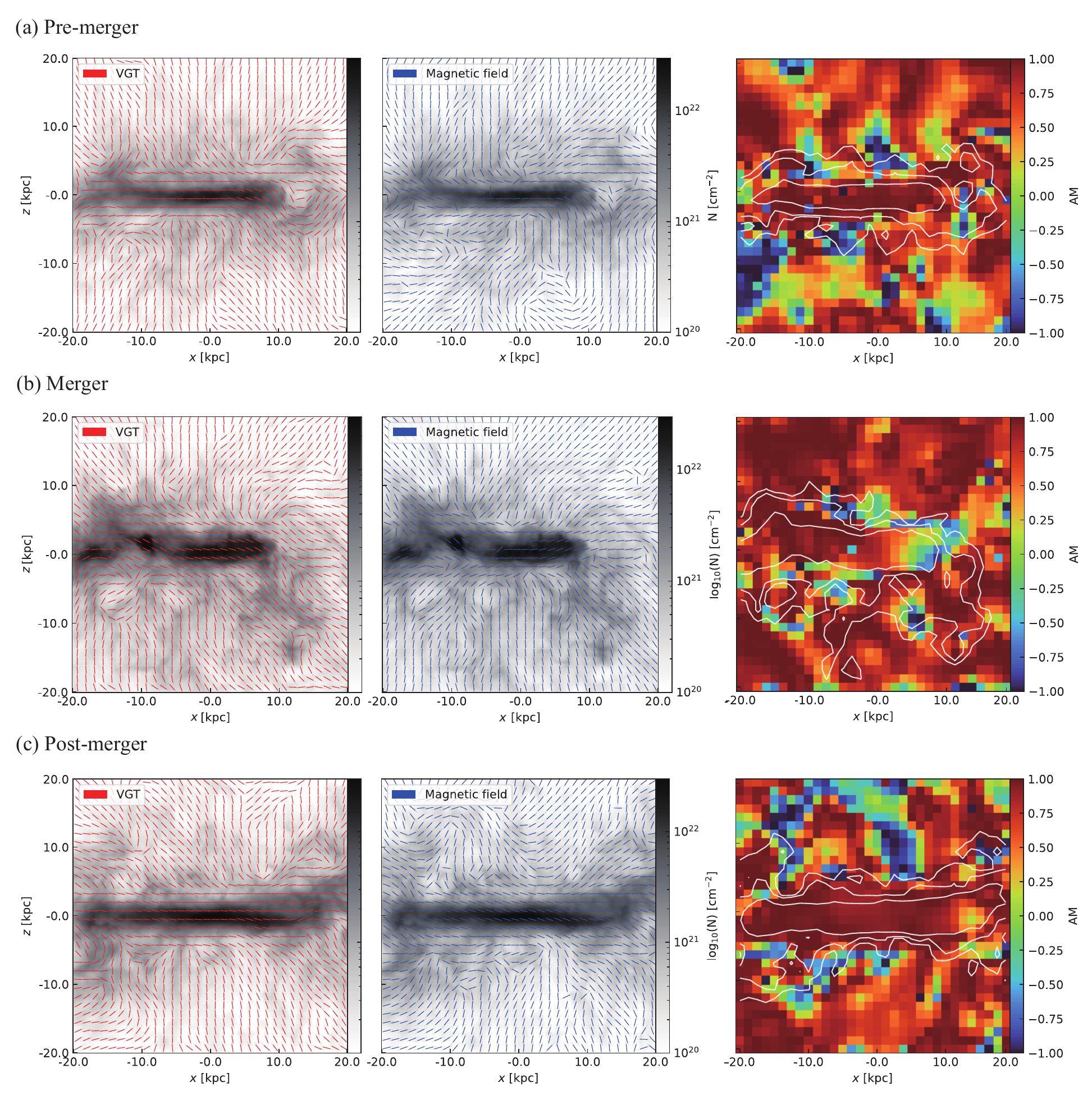}
        \caption{As Fig.~\ref{fig:vgt_060}, but for the edge-on cases.}
    \label{fig:vgt_edge}
\end{figure*}

\begin{figure}
\includegraphics[width=1.0\linewidth]{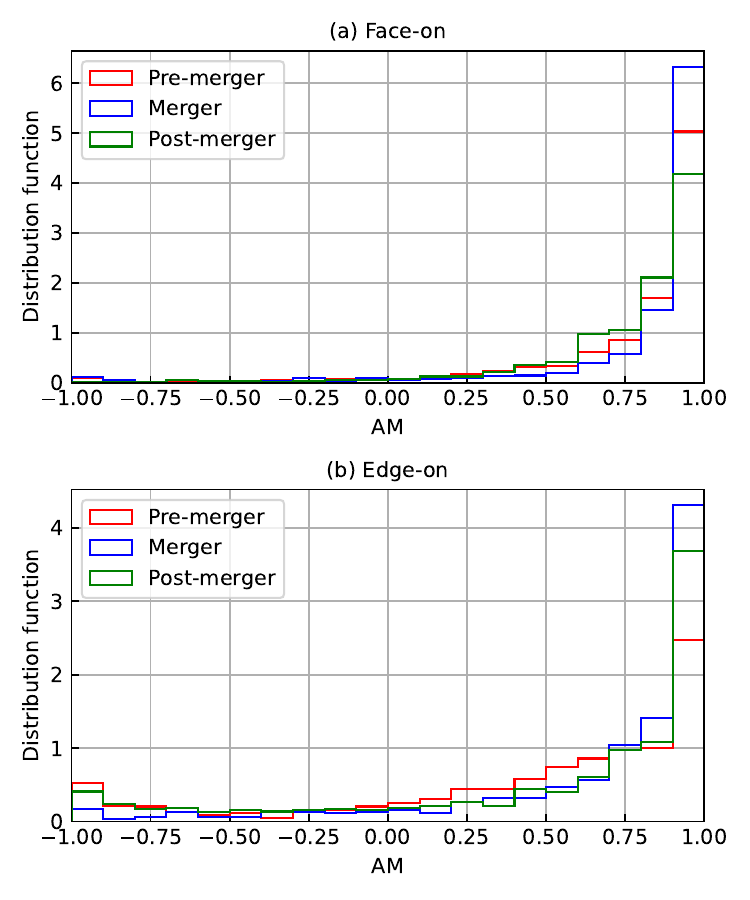}
        \caption{Distribution functions of alignment measure (AM) shown in Figs.~\ref{fig:vgt_060} and \ref{fig:vgt_edge}. A positive AM corresponds to parallel alignment between the two vectors, while a negative AM represents a perpendicular alignment, see Eq.~\ref{eq.am}.}
    \label{fig:am_hist}
\end{figure}
In the middle panel of Fig.~\ref{fig:Bvangle}, we show the histograms of the relative angle between the velocity gradient and the magnetic field in 3D space. 
The velocity gradient here is calculated at the pixel resolution of 78 pc, but this result holds for various scales tested. The histograms for all three stages peak at 90 degrees, implying that the velocity gradient is preferentially perpendicular to the magnetic field. This is an important validation of a key assumption of the VGT framework. Until this work, it had not been tested whether this relation held in non-idealized circumstances; i.e. in galaxies and galaxy mergers.

In Fig.~\ref{fig:Bvangle}, we show the amplitude of the velocity gradient as a function of separation. The slope of the gradients consistently curves, being steeper than the -2/3 expected for Kolmogorov turbulence at scales greater than 3 kpc but, for both pre- and merger stages, shallower than -1/2 for at scales smaller than 3 kpc. This may be due to contributions from other types of gradients, such as those induced by differential rotation. 
It is not completely clear what drives this, but we note that at this time the galaxy is at its most magnetized at this scale (see Figs.~\ref{fig:Bhist} and~\ref{fig:SF}).

Although there is some variation in the slope, it can be seen that in each merger stage, the amplitude increases monotonically as separation decreases. This aligns with expectations from MHD turbulence theory that gradients associated with turbulent velocity fluctuations dominate at small scales.   

\subsection{VGT: comparison with magnetic fields inferred from polarization}
In this section of our study, we focus on testing the efficacy of VGT when applied to the highly dynamic flows involved in galaxy mergers. This is key to checking that the VGT works in non-ideal circumstances. We include an analysis of the effects of orientation, by rotating our simulation boxes before generating the synthetic spectroscopic lines, covering face-on and edge-on orientations of galaxies.
\begin{figure*}
\centering
\includegraphics[width=0.75\linewidth]{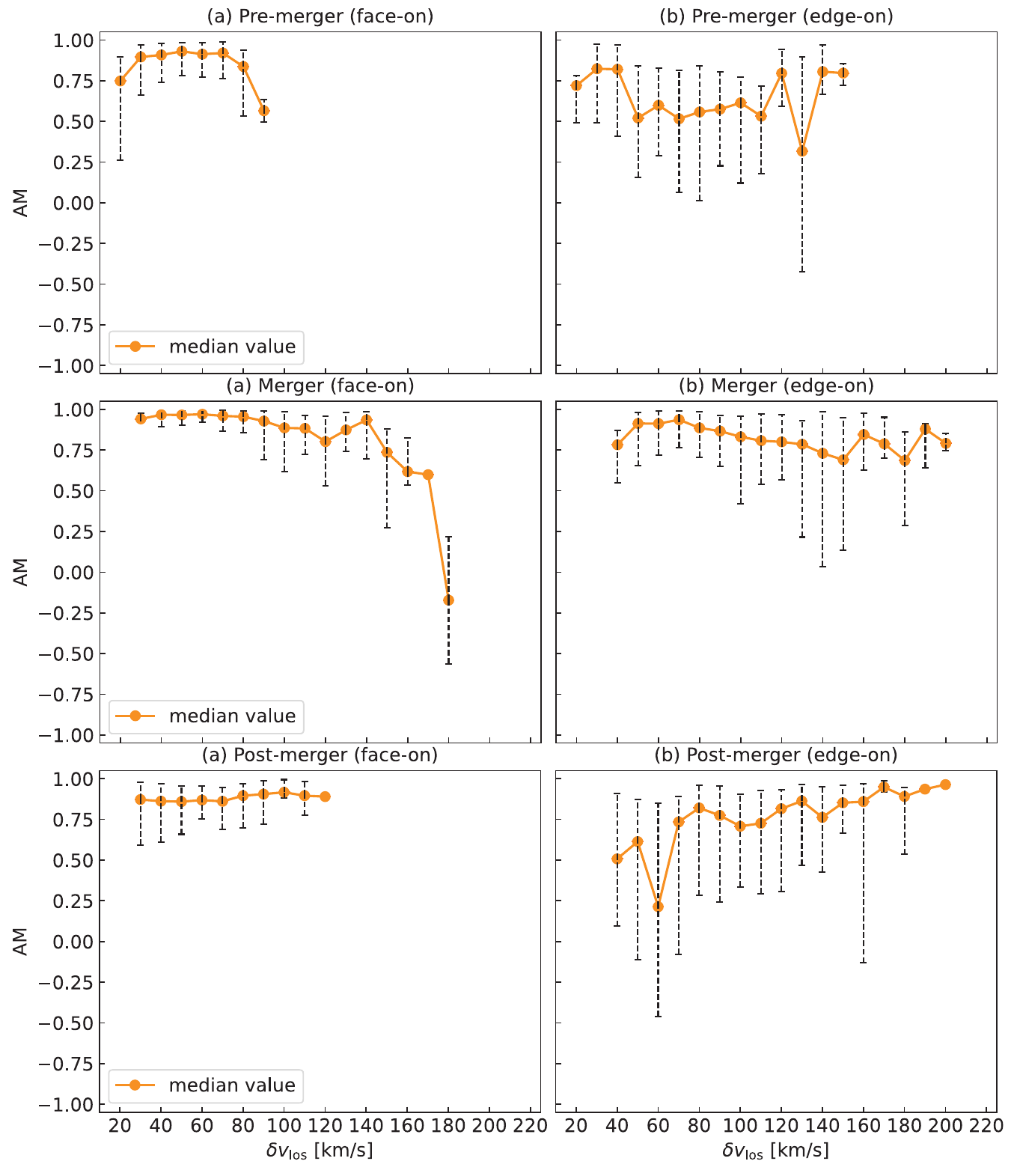}
        \caption{The change of AM as a function of the LOS velocity dispersion $\delta v_{\rm los}$. The AM is binned for every 10 km/s interval. The upper and lower black lines give the first (lower) and third quartiles (upper) and the orange line represents the median value. The lines for the first and third quartiles are not shown if they are too close to the median values, meaning only one or two samples fall into this range.}
    \label{fig:am_disv}
\end{figure*}

In Fig.~\ref{fig:vgt_060}, we present a comparative analysis of the magnetic field orientation in face-on galaxies, as mapped by the VGT and derived from synthetic polarization. We quantify the alignment between the VGT and polarization-derived magnetic fields with the aid of an AM map. We remind the reader that an AM score of 1 indicates perfect agreement, whilst an AM score of -1 indicates complete orthogonality between the two methods. In implementing the VGT, we employed the sub-block averaging method (see \S~\ref{sec:vgt}), which reduces the resolution of the magnetic fields. To facilitate a fair comparison, we correspondingly smooth the polarization-inferred magnetic field maps to match the VGT's resolution. This approach ensures that any differences observed are due to the intrinsic characteristics of the methods, rather than disparities in resolution.

For the face-on case, the magnetic fields inferred from VGT and polarization both exhibit a spiraling pattern that largely adheres to the structures of the galaxy at all times, as is expected from observations \citep[see, e.g.,][]{2015A&ARv..24....4B}. The magnetic fields measured by the VGT show a global agreement with those inferred from polarization. This demonstrates the general effectiveness of VGT in recovering the magnetic field structure. A more quantitative statement can be made with the alignment measure (AM) map shown on the right-hand side. 

Although alignment in Fig.~\ref{fig:vgt_060} is generally good, some local misalignments are also apparent. These may be attributed to the differing sensitivities of these techniques to specific physics. As discussed in \S\ref{sec:theory}, VGT, which is grounded in the dynamics of turbulent fluids, is sensitive to inflows and outflows \citep{2022ApJ...941...92H,2024MNRAS.530.1066L} as well as the anisotropy of velocity field. Examples of differences between the two methods can be seen, for instance, during the merger and post-merger, where areas of negative AM are evident in the center. We note that such negative AM is also observed in the center of the Centaurus galaxy merger \citep{2024arXiv240604242N}. Sites of negative AM will be apparent where shear velocity, merging flows, or nuclei activities are more pronounced.

In the edge-on scenarios shown in Fig.~\ref{fig:vgt_edge}, the magnetic field mostly aligns with the orientation of the galactic disk for both pre-merger and post-merger. During the merger, however, the configuration of the magnetic field becomes more complex. Similar to observations in the face-on case, the VGT generally agrees with the magnetic orientations inferred from polarization methods, particularly in the disk. Deviations are, however, evident further from the disk, where local misalignments occur. The significance of inflows and outflows likely becomes important away from the disk, as noted earlier in \S\ref{sec:structure-functions}. Velocity gradients in the presence of strong inflows or outflows are preferentially parallel to the magnetic field, rather than perpendicular, resulting in negative AM \citep{2020ApJ...897..123H,2021ApJ...912....2H,2024MNRAS.530.1066L}.

We quantify the degree of alignment between the two measures, by making histograms of the AM data shown in Figs.~\ref{fig:vgt_060}, and \ref{fig:vgt_edge}. These histograms are presented in Fig.~\ref{fig:am_hist}.
These histograms reveal that the AM values are primarily concentrated in the range of 0.5 to 1.0, with a notable peak at AM = 1. Alignment is especially good at the merging stage, which is also when gas flows are at their most turbulent. Edge-on orientations, however, tend to show a broader distribution of AM, with the distribution extending further into negative values. As noted previously, these negative AM values occur predominantly away from the disc, where inflows and outflows are more of an issue. Overall, though, across all three stages of the galactic merger process -- pre-merger, merger, and post-merger -- the AM values are predominantly positive, with the majority approaching an AM of 1. This indicates a substantial global agreement between the two methods in mapping magnetic field orientations. 

\subsection{Implication when VGT and polarization results are misaligned}
\label{subsec:misalignment}
Despite the overall good agreement between the two methods, we have identified that they are sometimes misaligned. This is particularly the case further away from the disc. There are several potential reasons for this. Firstly, VGT utilizes pseudo-Stokes parameters, as defined in Eq.~\ref{eq.QU}, to approximate polarization measurements. However, there are notable differences between these pseudo-Stokes parameters and the real Stokes parameters used in polarization measurements. For polarization observations, the Stokes parameters are calculated based on a density-weighted\footnote{Specifically, the relativistic electron density, in the case of synchrotron-derived polarization, and the dust density, in the case of dust-based polarization.} magnetic field \citep{2015A&A...576A.104P}. This approach inherently emphasizes regions of higher density, as they contribute more to the integrated polarization signal. However, the weighting in the pseudo-Stokes parameters used in VGT is based on the observed intensity in spectroscopic channels (see Eq.~\ref{eq.QU}). As discussed in \S~\ref{sec:theory}, the non-linear spectroscopic mapping from real spatial coordinates $(x,y,z)$ to velocity space $(x,y,v)$ means that weighting in the VGT is influenced more by the velocity field and its fluctuations. This difference in weighting mechanisms can result in VGT highlighting different parts of the magnetic field compared to polarization measurements; that is, regions with significant velocity fluctuations are likely to be more pronounced in VGT mappings \citep{2020ApJ...897..123H,2021ApJ...912....2H,2024MNRAS.530.1066L}. This sensitivity can be used to our advantage, however, as misalignment can then be potentially used to search for the presence of strong inflows and outflows. On the other hand, the Goldreich-Kylafis effect \citep{1981ApJ...243L..75G}, which uses spectroscopic observation to trace the magnetic field, may potentially test our hypothesis of the weighting differences in polarization and VGT.

Another plausible explanation for the observed misalignments in VGT measurements, as mentioned earlier in \S~\ref{sec:theory}, can be attributed to local variations in turbulent energy and magnetic field energy.  MHD turbulence exhibits differing characteristics based on the Alfv\'enic Mach number ($M_A$), which is the ratio of the turbulent injection velocity to the Alfv\'en velocity. In sub-Alfv\'enic conditions ($M_A < 1$), MHD turbulence is anisotropic, as the magnetic fields exert significant influence on the gas motions. The anisotropy, including anisotropy induced by non-turbulent motion, means that the velocity gradient is dominated by the component perpendicular to the magnetic field. Conversely, in super-Alfv\'enic conditions ($M_A > 1$), turbulence tends to be more isotropic due to the diminished back-reaction of the magnetic field on the turbulence\footnote{It is, however, important to note that the kinetic energy in turbulent motions typically follows a Kolmogorov cascade, characterized by $v_l \sim l^{1/3}$, where $v_l$ is the velocity at scale $l$. This implies that the influence of a magnetic back-reaction becomes more pronounced at smaller scales. Eventually, at the scale $l_{\rm A} = L_{\rm inj}M_{\rm A}^{-3}$, where $L_{\rm inj}$ is the injection scale of the turbulence, the turbulent velocity equals the Alfv\'en velocity, leading to a transition from super-Alfv\'enic to sub-Alfv\'enic turbulence \citep{2006ApJ...645L..25L}.  This means turbulence is still anisotropic below $l_{\rm A}$, providing another resolution requirement.}. At this point, the velocity gradient is not expected to be perpendicular to the magnetic field. Although Fig.~\ref{fig:Bvangle} shows the velocity field is anisotropic globally up to 3~kpc around, the local turbulence and fluid properties differ from the global one. Therefore, at a given resolution in observations or simulations, the velocity field could be locally isotropic. This variation in isotropy can lead to misalignments in the VGT mappings.

\subsection{The impact of velocity dispersion}
To explore the effect of super-Alfv\'enic turbulence, we study the correlation between AM and the LOS velocity dispersion $\delta v_{\rm los}$, as derived from spectroscopic data. The dispersion is defined as follows:
\begin{equation}
\begin{aligned}
        \delta v_{\rm los}&= \sqrt{\frac{\int p(x,y,v_{\rm los})(v_{\rm los}-\langle v_{\rm los}\rangle)^2dv_{\rm los}}{\int p(x,y,v_{\rm los})dv_{\rm los}}},\\
        \langle v_{\rm los}\rangle&=\frac{\int  p(x,y,v_{\rm los})v_{\rm los}dv_{\rm los}}{\int p(x,y,v_{\rm los})dv_{\rm los}},
\end{aligned}
\end{equation}
where $p(x,y,v_{\rm los})$ represents the intensity on the spectroscopic channel. A large dispersion indicates more significant turbulent velocity at the injection scale, which indicates that the contribution from local super-Alfv\'enic turbulence is likely larger. Advantageously, $\delta v_{\rm los}$ can be obtained from observations, allowing us to test the results. 

We show the correlation between $\delta v_{\rm los}$ and AM in Fig.~\ref{fig:am_disv}. From the face-on view, the pre-merger and merger stages exhibit similar trends, showing a decreasing AM with increasing $\delta v_{\rm los}$. This suggests that a large local velocity dispersion or super-Alfv\'enic condition could be one of the reasons for the observed misalignment between the VGT and polarization measurements. The AM, however, is more stable at the post-merger stage. In this stage, it is likely that the extra magnetic field amplification has reduced the amount of super-Alfv\'enic turbulence.

Conversely, in the edge-on view, there is no consistent increase or decrease of the AM with $\delta v_{\rm los}$. The edge-on perspective includes more variation in dynamical processes than the face-on cases, including, for example, inflows and outflows. These may override the effects of super-Alfv\'enic turbulence.

\section{Discussion}
\label{sec:dis}
\subsection{Anisotropic velocity fluctuations}
In this study, we observe that galactic mergers lead to a substantial increase in both velocity and density fluctuations, as well as an amplification of magnetic fields across all scales. This suggests that mergers play a significant role in driving turbulence, with this impact being continued to be felt long after coalescence. We also find that the slopes for the velocity structure functions are close to 1/2 up to approximately 0.3~kpc (see Fig.~\ref{fig:SF}). Beyond 3~kpc the slope becomes shallower, however, largely because of dominant large-scale galactic flows and motions occurring outside the ISM. On these scales, powerful AGN-driven outflows originating near the galactic center lead to magnetic draping, which likely contributes to a perpendicular alignment between velocity gradients and the magnetic fields. Star-formation-driven outflows in the outer disk create additional large-scale ``fountain''-type motions, some of which feed into a turbulent cascade. Cold clumps from the circumgalactic medium falling onto the disk further drape the circumgalactic magnetic field, potentially contributing to the perpendicular gradient–field alignment. Gas accretion at the disk outskirts may also contribute to the large-scale velocity fluctuations. 


By decomposing the velocity fluctuations into components parallel and perpendicular to local magnetic fields, we find that perpendicular velocity fluctuations are more significant than parallel ones, confirming that velocity fields in galaxies are predominantly anisotropic. However, the ratio of perpendicular to parallel components was observed to be nearly constant up to a scale of around 3~kpc (as shown in Fig.~\ref{fig:Bvangle}), while the ratio is expected to be scale dependent and magnetization dependent \citep{2021ApJ...911...37H}. Local isotropic super-Alfv\'enic turbulence or other non-turbulent velocities might affect the ratio. Further decomposition into turbulent and non-turbulent components may be necessary to retrieve these dependencies.

\subsection{VGT: implications for the required resolution in observational studies}
A key aspect of the VGT's underlying theory involves the scaling of turbulent velocity fluctuations. As indicated in Eqs.~\ref{eq.lv99}, ~\ref{eq.v}, and ~\ref{eq.vg}, the fluctuations should decrease at smaller scales, whilst the amplitude of the turbulent velocity gradient becomes more pronounced. This behavior contrasts with shear velocity gradients, which typically do not exhibit a similar increase in gradient at smaller scales. Hence, at these smaller scales, the turbulent velocity gradient is expected to dominate over shear velocity effects. Earlier studies assumed that the turbulent velocity gradient starts to dominate from around 0.1~kpc \citep{2022ApJ...941...92H,2023MNRAS.519.1068L,2024ApJ...967...18Z,2024arXiv240604242N}. This would necessitate high-resolution spectroscopic observations in order to resolve these scales. However, our analysis, as shown in Fig.~\ref{fig:SF}, shows that the velocity structure functions can maintain a slope of 1/2 up to approximately 0.3~kpc and that anisotropy is apparent up to 3~kpc. This finding extends the scale of turbulent velocity gradient dominance to 0.3~kpc and the scale of the perpendicularity between velocity gradient and magnetic fields to 3~kpc, a significant increase from the previously assumed scale. Consequently, the resolution requirement for observational studies utilizing VGT is substantially less stringent than initially thought. 

\section{Summary}
\label{sec:con}
The Velocity Gradient Technique (VGT; \citealt{GL17,LY18a,HYL18}) offers a novel method for mapping magnetic fields in external galaxies \citep{2022ApJ...941...92H,2023MNRAS.519.1068L}. Central to VGT’s methodology is the assumption that velocity fields in galaxies are anisotropic. Although VGT has been validated using ideal MHD turbulence simulations and observations of quiescent galaxies, it has yet to be tested in the context of galaxy mergers, where velocity fields are substantially more complex. To address this, we analyzed a zoom-in cosmological simulation of a galaxy merger using AREPO \citep{2021MNRAS.506..229W}, analyzing the pre-merger, merger, and post-merger stages. Our major findings are as follows:
\begin{enumerate}
    \item  We confirm that velocity fluctuations in galaxy mergers are distinctly anisotropic, with a more pronounced component perpendicular to the local magnetic field. The velocity gradient tends to be preferentially perpendicular to the magnetic field, with its amplitude increasing at smaller scales. 
    \item Our application of VGT across different merger stages reveals a statistically significant global agreement with magnetic fields as derived from polarization-like mock observations. This finding validates the effectiveness of VGT in tracing magnetic fields in different galaxy evolution stages and conditions.
    \item Throughout the galaxy merger, we observed a substantial intensification of velocity, magnetic field, and density fluctuations across all scales. The post-merger values for velocity and magnetic field structure functions were systematically higher than their pre-merger states \citep[see also][]{2021MNRAS.506..229W}.
    \item The scaling of velocity fluctuations exhibits a slope close to 1/2 up to a scale of approximately 0.3~kpc, before shifting to a steeper trend between 0.6 and 3~kpc, and to a shallower trend at larger scales. 
    The scaling of magnetic field and density fluctuations at scales smaller than 1.0 kpc also align closely with a 1/2 slope. 
\end{enumerate}

\begin{acknowledgments}
Y.H. acknowledges the support for this work provided by NASA through the NASA Hubble Fellowship grant No. HST-HF2-51557.001 awarded by the Space Telescope Science Institute, which is operated by the Association of Universities for Research in Astronomy, Incorporated, under NASA contract NAS5-26555. A.L. acknowledges the support of NSF grants AST 2307840, and ALMA SOSPADA-016. J.W. acknowledges the support of the German Science Foundation (DFG) under grant 444932369. C.P. acknowledges the support of the European Research Council under ERC-AdG grant PICOGAL-101019746. Financial support for this work was provided by NASA through award 09\_0231 issued by the Universities Space Research Association, Inc. (USRA). This work used SDSC Expanse CPU at SDSC through allocations PHY230032, PHY230033, PHY230091, and PHY230105 from the Advanced Cyberinfrastructure Coordination Ecosystem: Services \& Support (ACCESS) program, which is supported by National Science Foundation grants \#2138259, \#2138286, \#2138307, \#2137603, and \#2138296.
\end{acknowledgments}

%


\vspace{30mm}
\emph{Software:} \texttt{yt} \citep{Turk2011}, 
\texttt{Paicos} \citep{Berlok_Paicos_A_Python_2024},
\texttt{astropy} \citep{astropy}, \texttt{matplotlib} \citep{Hunter2007} and \texttt{numpy} \citep{numpy}.



\bibliography{example}{}
\bibliographystyle{aasjournal}



\end{document}